\documentclass[11pt,letterpaper,usenames,dvipsnames]{article}
\usepackage{jheppub}

\allowdisplaybreaks[2]  
\usepackage{pifont}
\newcommand{\cmark}{\text{\ding{51}}}
\usepackage{bm} 
\usepackage{dsfont} 

\usepackage{float}

\usepackage{tikz}
\usetikzlibrary{arrows,snakes,shapes.arrows,decorations.markings}
     \tikzset{>=triangle 90}
     \tikzstyle{gr}=[draw,circle,green!50!black,fill=green!50!black,scale=.6]
     \tikzstyle{Bl}=[draw,circle,blue,scale=.6]
     \tikzstyle{R}=[draw,circle,fill=red,scale=.6]
     \tikzstyle{bl}=[draw,circle,fill=black,scale=.35]
     \tikzstyle{bbc}=[draw,circle,fill=black,scale=.75]
     \tikzstyle{bbcs}=[draw,circle,fill=black,scale=.5]
     \tikzstyle{rc}=[circle,fill=red,scale=.6]
     \tikzstyle{wc}=[draw,circle,scale=.75]

\usepackage{array} 
\usepackage{multirow} 
\usepackage{colortbl} 
\usepackage{stfloats}  

\usepackage{xcolor}   

\def\ccb{\cellcolor{blue!07}}
\def\ccw{\cellcolor{white}}

\usepackage{url} 

\newcommand{\beq}{\begin{equation}}
\newcommand{\eeq}{\end{equation}}
\def\del{{\partial}}
\def\bar{\overline}
\def\til{\widetilde}
\def\hat{\widehat}
 
\def\vev#1{{\langle{#1}\rangle}} 

\def\^{\wedge}

\def\U{\mathop{\rm u}}
\def\SU{\mathop{\rm su}}

\def\SO{\mathop{\rm so}}

\def\GL{\mathop{\rm GL}}

\def\C{\mathbb{C}} 
\def\H{\mathbb{H}}

\def\R{\mathbb{R}} 
\def\Z{\mathbb{Z}} 
\def\ff{{\mathfrak f}}
\def\gf{{\mathfrak g}}

\def\tf{{\mathfrak t}}
\def\be{{\bf e}}

\def\bm{{\bf m}}

\def\bn{{\bf n}}

\def\br{{\bf r}}

\def\cH{{\mathcal H}}

\def\cM{{\mathcal M}}
\def\cN{{\mathcal N}}

\def\cS{{\mathcal S}}

\def\a{{\alpha}}

\def\ba{{\boldsymbol\a}}
\def\b{{\beta}}

\def\g{{\gamma}}

\def\G{{\Gamma}}

\def\D{{\Delta}}

\def\l{{\lambda}}

\def\m{{\mu}}
\def\bmu{{\boldsymbol\m}}

\def\r{{\rho}}

\def\f{{\phi}}

\def\w{{\omega}}

\def\ssqt{{\scriptscriptstyle\sqrt2}}

\title{Expanding the landscape of $\cN=2$ rank 1 SCFTs}
\author{Philip C. Argyres,}
\author{Matteo Lotito,}
\author{Yongchao L\"u}
\author{and Mario Martone}
\affiliation{University of Cincinnati,
Physics Department, PO Box 210011, Cincinnati OH 45221}
\emailAdd{philip.argyres@gmail.com}
\emailAdd{lotitomo@mail.uc.edu}
\emailAdd{lychaoaa@gmail.com}
\emailAdd{martonmo@ucmail.uc.edu}

\abstract{
We refine our previous proposal \cite{Argyres:2015ffa, Argyres:2015gha, Argyres:2015ccharges} for systematically classifying 4d rank-1 $\cN=2$ SCFTs by constructing their possible Coulomb branch geometries.  Four new recently discussed rank-1 theories \cite{Garcia-Etxebarria:2015wns,Chacaltana:2016shw}, including novel $\cN=3$ SCFTs, sit beautifully in our refined classification framework.  By arguing for the consistency of their RG flows we can make a strong case for the existence of at least four additional rank-1 SCFTs, nearly doubling the number of known rank-1 SCFTs.  

The refinement consists of relaxing the assumption that the flavor symmetries of the SCFTs have no discrete factors.   This results in an enlarged (but finite) set of possible rank-1 SCFTs.  Their existence can be further constrained using consistency of their central charges and RG flows.
}
 
\begin{document}
\maketitle

\section{Introduction
\label{sec1}}

In a series of three papers \cite{Argyres:2015ffa, Argyres:2015gha, Argyres:2015ccharges} we have outlined a strategy for a systematic classification of 4d $\cN=2$ SCFTs and carried it out for the regular rank-1 case.  This is the case where the Coulomb branch (CB) is one complex-dimensional, with parameter $u\in \C$.  In \cite{Argyres:2015gha} we argued the case for the existence of up to 11 rank-1 $\cN=2$ SCFTs (see table 1 in \cite{Argyres:2015gha}) in addition to the 11 already known in the literature \cite{Seiberg:1994aj,Argyres:1995xn,Minahan:1996fg,Minahan:1996cj,Argyres:2007cn,Argyres:2007tq}. 

Although this represents a dramatic enlargement of the set of possible rank-1 SCFTs, we will argue here that four of these theories have already been constructed using string theory or $\cS$-class techniques.  These are:
\begin{itemize}
\item Three new rank-1 SCFTs \cite{Garcia-Etxebarria:2015wns} with $\D(u) = \ell$, $\ell\in\{3,4,6\}$ and an abelian flavor symmetry $\ff=\U(1)$.  If the single allowed mass deformation is turned off, these theories enjoy an $\cN=3$ enlarged supersymmetry.
\item A new rank-1 SCFT \cite{Chacaltana:2016shw} with $\D(u)=6$ and $\ff=A_3$.  The $\cS$-class curve \cite{Gaiotto:2009we} for this theory only makes its $A_1\oplus A_1\subset A_3$ mass deformations explicit.
\end{itemize}
The flavor symmetries of these theories do not all match the ones predicted in \cite{Argyres:2015gha}.  The reason for the mismatch is the assumption made in \cite{Argyres:2015gha} that the discrete symmetry group, $\G$, of the CB geometry should be interpreted as the Weyl group of the flavor symmetry: $\G=\text{Weyl}(\ff)$. By weakening this assumption so that only a subgroup, $\G'\subset \G$, is the Weyl group of the flavor symmetry Lie algebra, $\G'=\text{Weyl}(\ff')$, other choices of the flavor symmetries become consistent.  As we will explain in more detail below, if there is a complementary subgroup\footnote{$\G'$ and $\G''$ are complementary if $\G'\G''=\G$ and $\G'\cap \G''=\{1\}$.} $\G''\subset\G$ which acts as an outer automorphism of $\ff'$, then it is consistent to interpret the flavor symmetry of the CB geometry as $\G''\ltimes\ff'$ instead of $\ff$.  It turns out that the possibilities for $\G''$ and $\ff'$ are quite limited, so only a few additional flavor assignments are allowed for each geometry reported in \cite{Argyres:2015gha}.

Consistency of the RG flows among these theories puts additional constraints on the existence of SCFTs with these flavor symmetries.  These constraints allow us to rule out certain flavor assignments as inconsistent.  Conversely, RG flows from the four newly constructed theories mentioned above allow us to deduce the existence of at least four additional SCFTs and determine their flavor symmetries and central charges.

\begin{table}[ht]
\centering
$\begin{array}{|c|c|c|c|c|c|c|c|}
\hline
\multicolumn{8}{|c|}{
\multirow{2}{*}{\text{\large\bf Some rank 1 $\cN=2$ SCFTs}}}\\
\multicolumn{8}{|c|}{}\\
\hline\hline
\text{Kodaira}  & \text{deformation} & \text{flavor} 
& \multicolumn{3}{c|}{\text{central charges}}
& \multicolumn{2}{c|}{\text{Higgs branches}} \\[-.5mm] 
\ \text{singularity}\ \,  & \text{pattern} & \text{symmetry} &k_\ff & 
12\cdot c &\ 24\cdot a\ \, &\quad h_1\quad\, & h_0 \\ 
\hline
\multirow{6}{*}{$II^*$}   
&\{{I_1}^{10}\} &E_8  &12 &62 &95 &0 &29 \\ 
&\{{I_1}^6,I_4\} &C_5 &7  &49 &82 &5 &16 \\ 
&\ccb \{{I_1}^{3},I_1^*\} &\ccb A_3\rtimes\Z_2 
&\ccb 14 &\ccb 42 &\ccb 75 &\ccb 4 &\ccb 9 \\ 
&\ccb\{{I_1}^{2},IV^*_{Q=1}\} &\ccb A_2\rtimes\Z_2 
&\ccb 14 &\ccb 38 &\ccb 71 &\ccb3 &\ccb? \\
\rowcolor{blue!07}\ccw &\{I_2,IV^*_{Q=\ssqt}\}& \U(1)\rtimes\Z_2       
&? &33 &66&1 &1 \\
\rowcolor{blue!07}\ccw &\{I_1,III^*\} &\U(1)\rtimes\Z_2 
& ? & 33  & 66 &1 &1\\
\hline 
\multirow{5}{*}{$III^*$}  
&\{{I_1}^9\} &E_7 &8 &38 &59 &0 &17  \\ 
&\{{I_1}^5,I_4\} &\ C_3\oplus A_1\ \,
&5\oplus8 &29 &50 &\ 3\ \, &8 \\ 
&\ccb  \{{I_1}^2,I_1^*\} &\ccb \ A_1\oplus(\U(1)\rtimes\Z_2)\ \, 
&\ccb \ 10\,\oplus\, ?\ \, &\ccb 24 &\ccb 45 &\ccb 2 &\ccb ? \\ 
&\ccb\{I_1,IV^*_{Q=1}\} &\ccb\U(1)\rtimes\Z_2
& \ccb ?  &\ccb21 &\ccb42 &\ccb1 &\ccb1  \\
\rowcolor{blue!07}\ccw & III^*&\varnothing
& - &18 & 39 &0 &0  \\
\hline
\multirow{5}{*}{$IV^*$} 
&\{{I_1}^8\} &E_6 &6 &26 &41 &0 &11 \\ 
&\{{I_1}^4,I_4\} &C_2\oplus\U(1) &4\oplus{?}
&19 &34 &2 &4 \\ 
&\{I_1,I^*_1\}\ccb &\ccb\U(1) 
&\ccb ? &\ccb15 &\ccb30 &\ccb1 &\ccb1 \\ 
\rowcolor{blue!07}\ccw &IV^*_{Q=\ssqt} &\varnothing
&-&14 & 29 &0 &0  \\
\rowcolor{blue!07}\ccw &IV^*_{Q=1} &\varnothing
&- &25/2 & 55/2 &0 &0  \\[1.5mm]
\hline
\multirow{2}{*}{$I^*_0$} 
&\{{I_1}^6\} &D_4 &4 &14 &23 &0 &5 \\ 
&\ \{{I_1}^2,I_4\} \simeq \{{I_2}^3\}\ \,&A_1 
&3 &9 &18 &1 &1 \\ 
\hline
IV 
&\{{I_1}^4\} &A_2 &3 &8 &14 &0 &2 \\ 
\hline
III
&\{{I_1}^3\} &A_1 &8/3 &6 &11 &0 &1 \\ 
\hline
II 
&\{{I_1}^2\} &\varnothing 
&- &\ 22/5\ \, &\ 43/5\ \,&0 &0 \\ 
\hline
\end{array}$
\caption{\label{tab1} 
Predicted properties of some regular rank-1 $\cN=2$ SCFTs are listed.  ``Kodaira singularity" refers to the Kodaira type of the scale invariant CB geometry, and ``deformation pattern" lists the resulting singularity types under a generic relevant deformation of the SCFT.  $h_0$ and $h_1$ refer to the quaternionic dimensions of the Higgs branch and of the enhanced Coulomb branch (ECB) fibers, respectively.  There are question marks where there is not enough information from the CB geometry to usefully constrain an entry.  Theories supported by new evidence are shaded blue and un-shaded rows are for already established SCFTs.}
\end{table}

We partially summarize our results for the RG-flow consistent rank-1 theories in table \ref{tab1}.  Theories for which new evidence for their existence is presented are shaded blue in the table.\footnote{There are two theories reported in table \ref{tab1} (the $II^*\to\{I_1,III^*\}$ and $II^*\to\{I_2,IV^*_{Q=\ssqt}\}$ theories) which could be identified with a $II^*$ $\cN=3$ SCFT.  It is not clear whether only one or both should be identified as $\cN=3$ theories.  
The $Q=\sqrt2$ subscript means that BPS states on the CB of the $IV^*_{Q=\ssqt}$ have electric and magnetic charges which are multiples of $\sqrt{2}$. The reason we emphasize this is because there is a second possible frozen $IV^*$ SCFT, $IV^*_{Q=1}$, whose BPS states are quantized in units of $1$ (i.e., integers); see \cite{Argyres:2015ffa, Argyres:2015gha}.} We emphasize, however, that table \ref{tab1} only lists a fraction (about 2/5) of the total number of possible SCFTs with internally consistent RG flows. We summarize the RG flow constraints on all the possible SCFTs in figures \ref{I4}-\ref{IVs} in section \ref{sec4}. In particular there are additional theories, including one with $F_4$ flavor group, which fall in the $I_0^*$ series \cite{Argyres:2015gha} and will be discussed elsewhere \cite{Argyres:2016I0s}. 

The RG flow constraints can be organized in terms of three categories of flows, which we call \emph{matching}, \emph{compatible}, and \emph{unphysical} flows.  (These flows correspond to green, blue, and red arrows, respectively, in the figures in section \ref{sec4}.)  Matching flows are ones under which precisely the subgroup of the UV flavor symmetry which is not broken by the relevant operator intiating the flow is realized as flavor symmetries of the IR SCFTs on the CB.  Compatible flows are ones where the IR flavor group is accidentally enlarged, but its rank is not.  All other cases are unphysical flows.  These latter violate the safely irrelevant conjecture of \cite{Argyres:2015ffa}, and, as discussed in \cite{Argyres:2015gha} and in section \ref{sec4} below, do not have a consistent field theory interpretation.

We use these flows to label CB geometries together with a flavor symmetry assignment as \emph{good}, \emph{ugly}, or \emph{bad}.  (These theories are shown in green, blue, and red boxes, respectively, in the figures in section \ref{sec4}.)  Good theories are ones for which there exist matching flows for all relevant operators.  All the theories listed in table \ref{tab1} are good theories; however, there are additional good theories which are not shown there.  Ugly theories are ones for which at least one relevant operator induces a compatible flow, but no flows are unphysical.  We do not know of a first principles reason why such theories should not be allowed.  Finally, bad theories are those for which at least one relevant operator induces an unphysical flow.

The most interesting example of a good theory not shown in table \ref{tab1} is the $[II^*,G_2]$ theory.\footnote{We denote theories by $[K,\ff]$ where $K$ is the Kodaira type of their singularity and $\ff$ is their flavor symmetry.  The Kodaira types are correlated to dimensions of the CB parameters as $K=\{II^*, III^*, IV^*, I_n^*, IV, III, II, I_n\} \leftrightarrow \D(u)= \{6, 4, 3, 2, 3/2, 4/3, 6/5, 1\}$ respectively.}  This is an alternative flavor assignment to the $[II^*,A_2\rtimes\Z_2]$ theory shown in the fourth line of table \ref{tab1}.  As shown in figure \ref{IVs}, the $[II^*,G_2]$ does not flow to a $[III^*,\U(1)\rtimes\Z_2]$ but to a $[III^*,A_1]$ which is an alternative good interpretation of the former.  The other good theories not shown in table \ref{tab1} are similar alternative $A_1$ interpretations of the $\U(1)\rtimes\Z_2$ theories.  

We should note that the interpretation --- discussed at length in \cite{Argyres:2015gha} --- of the frozen $I_1^*$ and $I_0^*$ singularities as weakly gauged rank-0 SCFTs is not considered here.  For the $I_1^*$ singularity we focus only on the more conservative interpretation of this singularity as the lagrangian $\SU(2)$ gauge theory with a single half-hypermultiplet in the spin-3/2 representation.  Thus, in particular, the $I_1^*$ series shown in table \ref{tab1} --- i.e., the $[II^*,A_3\rtimes\Z_2]$, $[III^*,A_1\oplus(\U(1)\rtimes\Z_2)]$, and $[IV^*,\U(1)]$ theories --- are analyzed assuming this interpretation of the theory along each flow.  The $I_0^*$ case is more interesting,
and will be discussed separately \cite{Argyres:2016I0s}.



The paper is organized as follows.  Section \ref{sec2} describes the allowed flavor symmetry identifications of CB geometries.  Section \ref{sec3} analyzes the CB geometries of the new rank-1 theories constructed in \cite{Garcia-Etxebarria:2015wns, Chacaltana:2016shw}.  In section \ref{sec3.3} we present evidence supporting their identification with the geometries shown in table \ref{tab1} by matching their curve discriminants, RG flows, central charges, and ECB fibers. The discussion of the determination of central charges, which uses a combination of the techniques of \cite{Shapere:2008zf} combined with constraints coming from S-dualities, RG flows, and the properties of $\cN=3$ theories \cite{Aharony:2015oyb}, is presented in more detail in \cite{Argyres:2015ccharges}.  The results we find for the central charges of the $\cN=3$ theories agree with those found in \cite{Nishinaka:2016hbw}.  Section \ref{sec4} discusses more broadly the RG flow consistency constraints on all possible flavor symmetry assignments of CB geometries.  We then conclude with some final remarks and open questions.

\section{Discrete parts of flavor symmetries
\label{sec2}}

In \cite{Argyres:2015gha} we classified potential rank-1 Coulomb branch geometries of SCFTs by constructing possible inequivalent regular special K\"ahler mass deformations of scale-invariant Kodaira singularities.  The list of such deformations is given in table 1 of \cite{Argyres:2015gha}, where we also identified the flavor symmetry algebra, $\ff$, of the SCFTs associated to each geometry.  We determined these flavor symmetries as follows.

A mass-deformed Coulomb branch (CB) geometry is invariant under a discrete symmetry group, $\G$, which acts linearly on the $r$ independent linear mass parameters.  If we denote the complex masses by $\bm\in\C^r$, then the action of $\g\in\G$ is given by an $r\times r$ real matrix representation,
\begin{align}\label{CSAact}
\g:\bm \mapsto \r(\g) \bm, \qquad
\r(\g) \in \GL(r, \R).
\end{align}
We will refer to $\G$ as the (discrete) symmetry of the deformed CB.  In \cite{Argyres:2015gha} we found that $\G$ is always isomorphic to a real crystallographic reflection group, so, in particular, we can choose a basis of $\C^r$ so that $\r \in O(r)$, the group of $r\times r$ orthogonal matrices.  This means that $\G$ can be interpreted as the Weyl group of a reductive Lie algebra, $F_\G$,  
\begin{align}\label{}
\G = \text{Weyl}(F_\G) \qquad \text{with} \quad
\text{rank}(F_\G) = r.
\end{align}
Then the real representation $\r$ of $\G$ on $\C^r$ is the action of Weyl$(F_\G)$ on $\tf_\C \simeq \C^r$, the complexified Cartan subalgebra of $F_\G$.   Since the complex masses transform in the adjoint of the flavor algebra of an $\cN=2$ SCFT, and can be rotated by a flavor transformation to lie in $\tf_\C$, and for generic values break the flavor symmetry as
\begin{align}\label{maxadj}
\ff \xrightarrow{\text{adj}} \text{Weyl}(\ff) \circ_\r \tf ,
\end{align}
it is natural to identify the flavor symmetry with $F_\G$,
\begin{align}\label{maxflavor}
\ff \simeq F_\G .
\end{align}
In \eqref{maxadj} we have denoted the linear action \eqref{CSAact} of Weyl$(\ff)$ on $\tf$ by $\circ_\r$.

Note that the identification \eqref{maxflavor} may fail to determine the flavor symmetry algebra because the simple Lie algebras of Dynkin type $B_r$ and $C_r$ share the same Weyl group, so such factors cannot be distinguished.  We will therefore often call them $BC_r$ factors in what follows.\footnote{As discussed in \cite{Argyres:2015gha}, it may happen that the poles of the SW 1-form fill out Weyl orbits in flavor weight space which lie in the root lattice of $B_r$ but not of $C_r$, or \emph{vice versa}, in which case the 1-form distinguishes between the two Lie algebras.  Also, RG flows from one theory to another with an ambiguous $BC_r$ flavor group may only be consistent for only one of $B_r$ or $C_r$, again determining the flavor algebra.  These mechanisms account for the flavor assignments of three entries in table 1 of \cite{Argyres:2015gha}:  $II^*\to\{{I_1}^6,I_4\}$ ($\ff=C_5$), $III^*\to\{{I_1}^5,I_4\}$ ($\ff=C_3\oplus A_1$), and $III^*\to\{{I_1}^3,I_0^*\}$ ($\ff=B_3$).}  

However, it is logically possible that only a subgroup, $\G' \subset \G$, is the flavor algebra Weyl group.  Thus the connected part of the flavor algebra, $\ff_\text{conn.}$, may instead be identified with
\begin{align}\label{}
\ff_\text{conn.} \simeq F_{\G'}, \qquad \text{where}\quad 
 \text{Weyl}(F_{\G'})\simeq\G' \subset\G
\qquad \text{and}\quad 
\text{rank}(F_{\G'}) =\text{rank}(F_{\G})=r.
\end{align}
The rank must stay the same, since it is the number of linearly independent mass parameters, which is fixed.  So we are looking for  subgroups $\G'\subset \G$ which are also Weyl groups of a rank-$r$ Lie algebra.  This requirement may be satisfied, if necessary, by adding $\U(1)$ factors in $F_{\G'}$ since Weyl$(\U(1))$ is trivial.

The elements of $\G$ not in $\G'$ generate a group of discrete symmetries of the CB geometry, so should be included as an additional discrete part of the flavor symmetry, $\ff$.  To be a symmetry of the theory they must act by automorphisms of $F_{\G'}$,
\begin{align}\label{submaxflavor}
\ff \simeq \G'' \ltimes F_{\G'}, \qquad
\G'' \subset \text{Aut}(F_{\G'}).
\end{align}
Upon turning on masses, this means that there must be two subgroups $\G', \G''\subset \G$ such that $\G'\cap\G''=\{1\}$ and $\G''\circ_\r (\G' \circ_\r \tf)= \G\circ_\r\tf$.  Since $\G$ acts faithfully on $\tf$ via $\r$, this means that $\G''\G'=\G$ (and thus $\G'\G''=\G$).  Furthermore, $\G''$ must a be group of outer automorphisms of the connected flavor algebra, Out$(F_{\G'})$, since, by construction, its elements are non-trivial maps of the Cartan subalgebra of $F_{\G'}$ to itself, and inner automorphisms that preserve the Cartan subalgebra as a set only act non-trivially on it as Weyl$(F_{\G'})=\G'$.

These conditions are quite restrictive since the set of outer automorphisms of reductive Lie algebras is small, namely
\begin{align}\label{}
\begin{array}{c|ccccccccccc}
\gf &
\ \U(1)\  & \ A_1\  & \ A_{r>1}\  & \ BC_r\  & \ D_4\  & \ D_{r>4}\  &\  E_6\  &\ E_7\  & \ E_8\  &\ F_4\ &\ G_2\ \\
\hline
\text{Out}(\gf) &
\Z_2 & - & \Z_2 & - & S_3 & \Z_2 & \Z_2 & - & - & - & -
\end{array}
\end{align}
and Out$(\gf^n) = S_n \ltimes \text{Out}(\gf)$ for $\gf$ semi-simple,\footnote{We abuse notation here by writing the $n$-fold direct sum, $\gf \oplus \cdots \oplus \gf$, multiplicatively as $ \gf^n$.} and Out$(\U(1)^n) = O(n,\R)$.
Using some facts about the Weyl groups of simple Lie algebras \cite{mckay1981tables},
\begin{align}\label{weylinfo}
\begin{array}{c|cc}
\gf & \ \text{Weyl}(\gf)\ & \ |\text{Weyl}(\gf)| \ \\
\hline
A_r & S_{r+1} & \ (r+1)!  \ \\
BC_r & \ S_r \ltimes \Z_2^r \ & 2^r r! \\
D_r & \ S_r \ltimes \Z_2^{r-1} \ & 2^{r-1} r! \\
G_2 & \ \Z_2 \ltimes S_3 \ & 2^2\cdot 3 \\
F_4 & \ S_3 \ltimes \text{Weyl}(D_4)\ & 2^7 \cdot 3^2 \\
E_6 & \ldots & \ 2^7\cdot 3^4\cdot 5 \ \\ 
E_7 & \ldots & \ 2^{10}\cdot 3^4\cdot 5 \cdot 7 \ \\
E_8 & \ldots & 2^{14}\cdot 3^5 \cdot 5^2 \cdot 7  
\end{array} 
\end{align}
and the fact that if $\G'\G''=\G$ and $\G'\cap\G''=\{1\}$ then $|\G'|\cdot |\G''| = |\G|$, it is not too hard to list the simple $\ff_r$ and reductive $\ff'_r$ of rank $r$ and discrete group $\G''\subset \text{Out}(\ff'_r)$ such that Weyl$(\ff_r) = \G'' \cdot \text{Weyl}(\ff'_r)$:
\begin{align}\label{misident}
\begin{array}{c|cc}
\ff_r & \ \G''\ & \ \ff'_r \ \\
\hline
\gf &\ \text{Weyl}(\gf)\ &\ \U(1)^r \\
\ BC_r\  & S_r & {A_1}^r \\
BC_r & \Z_2 & D_r \\
G_2 & \Z_2 & A_2 \\
F_4 & S_3 & D_4
\end{array}
\end{align}
This gives the list of possible misidentifications of the flavor symmetry that may have been made in \cite{Argyres:2015gha}: any theory with a flavor symmetry $\ff$ in the left-most column of \eqref{misident} can be re-interpreted as a theory with flavor symmetry $\G''\ltimes\ff'$ instead.

The first line in \eqref{misident} is the tautological case where any rank-$r$ Weyl group can be re-interpreted as a discrete symmetry acting on a theory with a $\U(1)^r$ symmetry.  The second and third lines are true for all positive $r$ using the Lie algebra identifications $BC_1=B_1=C_1 = A_1$, $BC_2 = B_2 = C_2$, $D_1 = \U(1)$, $D_2 = A_1 \oplus A_1$, and $D_3 = A_3$.  In particular, the $r=1,2,3$ cases of the third line in \eqref{misident} are equivalent to 
\begin{align}\label{misident2}
\text{Weyl}(A_1) &= \Z_2 \cdot \text{Weyl}(\U(1)),
\nonumber\\
\text{Weyl}(C_2) &= \Z_2 \cdot \text{Weyl}(A_1\oplus A_1),
\\
\text{Weyl}(BC_3) &= \Z_2 \cdot \text{Weyl}(A_3).
\nonumber
\end{align}

We have focused so far only on the symmetry of the SW curve.  But this symmetry extends to the SW 1-form as well.  Since the 1-forms constructed in \cite{Argyres:2015gha} were by design invariant under the full discrete symmetry group $\G$, they are \emph{a fortiori} invariant under any subgroup $\G'\subset\G$.

The rest of this paper will explore the consequences of these possible misidentifications on the classification of rank-1 SCFTs.  In particular,  in \cite{Argyres:2015gha} we made just such a misidentification in the CB geometries of three of the four rank-1 SCFTs constructed in \cite{Garcia-Etxebarria:2015wns,Chacaltana:2016shw} as we will now argue.

\section{CB geometries of new SCFTs
\label{sec3}}

We start with the rank-1 SCFT found in \cite{Chacaltana:2016shw}.

\subsection{The $\ff=A_3$ theory with $\D(u)=6$
\label{sec3.1}}

In \cite{Chacaltana:2016shw}, by looking at the $D_4$ 6d $(2,0)$ SCFT twisted and compactified on 3-punctured spheres (``fixtures") with puncture boundary conditions including $\Z_3$ outer-automorphism twists of $D_4$, Chacaltana, Distler, and Trimm find a new rank-1 SCFT with central charges $a=25/8$, $c=7/2$, CB parameter of dimension $\D(u)=6$, and flavor algebra $\ff=A_3$ with flavor central charge $k=14$.

Since $\D(u)=6$, the curve describing the mass deformations of this SCFT must be a deformation of the type $II^*$ Kodaira singularity.  Among those listed in table 1 of \cite{Argyres:2015gha}, there is only one with a rank-3 flavor symmetry, namely the $[II^*,BC_3]$ curve: the $II^* \to \{ I_1^3, I_1^*\}$ deformation with flavor symmetry algebra identified as $BC_3$.   As the third entry in \eqref{misident2} shows, this can instead be reinterpreted as having flavor symmetry
\begin{align}\label{CDTflavor}
\ff = \Z_2 \ltimes A_3,
\end{align}
where the $\Z_2$  is the outer automorphism which acts on $A_3$ by conjugation.  So this identification predicts that the SCFT's flavor symmetry has the discrete $\Z_2$ factor shown in \eqref{CDTflavor}.

One can identify this discrete $\Z_2$ flavor factor from the class-$\cS$ construction of the SCFT in \cite{Chacaltana:2016shw}.  In that construction, the manifest $A_1\oplus A_1\subset A_3$ flavor symmetry comes from two $\Z_3$-twisted punctures each carrying an $A_1$ global symmetry and opposite $\Z_3$ twists.  The $\Z_2$ outer automorphism of $\Z_3$ interchanges these two punctures but leaves the fixture unchanged.  This $\Z_2$ thus interchanges the two $A_1$ factors in the manifest flavor algebra which is compatible with the action of the complex conjugation outer automorphism of $A_3$.\footnote{We thank J. Distler for pointing this out to us.}  

The $[II^*,BC_3]$ curve reported in equations (A.25)--(A.27)  of \cite{Argyres:2015gha} is written in terms of the $(m_1, m_2, m_3)$ linear masses associated with the flavor group $BC_3$.  Weyl$(BC_3) \simeq  {\Z_2}^3 \rtimes S_3$ acts on the $m_a$ by independent sign flips and permutations.  Abstractly, Weyl$(A_3)\simeq S_4$ is a subgroup of Weyl$(BC_3)$ since ${\Z_2}^3 \rtimes S_3 \simeq \Z_2 \rtimes ({\Z_2}^2 \rtimes S_3) \simeq \Z_2 \rtimes S_4$.  The Weyl$(A_3)$ action on the masses is usually written using a basis of four masses, $(\m_1,\m_2,\m_3,\m_4)$, satisfying the relation $\sum_i \m_i=0$, corresponding to the eigenvalues of an $SU(4)$ matrix.  Then Weyl$(A_3)$ acts by permutations of the $\m_i$.  An explicit relation between the $BC_3$ basis $m_a$ and the $A_3$ basis $\m_i$ is given by
\begin{align}\label{BC3toA3}
\m_1 &=-m_1+m_2+m_3 ,\nonumber\\
\m_2 &=+m_1-m_2+m_3 ,\nonumber\\
\m_3 &=+m_1+m_2-m_3 ,\\
\m_4 &=-m_1-m_2-m_3 ,\nonumber
\end{align}
whose inverse is $m_3 = \frac12 (\m_1+\m_2)$ and cyclic permutations of the $123$ indices.  Thus we propose that the curve and one-form of the $A_3$ SCFT of \cite{Chacaltana:2016shw} is the same as that of the $BC_3$ geometry described in Appendix A.1.4 of \cite{Argyres:2015gha} but with the $m_a$ everywhere substituted with the $\m_i$ via \eqref{BC3toA3}.

For later use, it will be useful to identify the action of the $\Z_2$ outer automorphism of $A_3$ in terms of both the $A_3$ and the $BC_3$ mass bases.  A standard basis of simple roots of $A_3$ is one in which the $(\ba_1)\!\!-\!\!(\ba_2)\!\!-\!\!(\ba_3)$ Dynkin diagram has
\begin{align}\label{}
\ba_1 &= \be_1 - \be_2,&
\ba_2 &= \be_2 - \be_3,&
\ba_3 &= \be_3 - \be_4,
\end{align}
with the $\be_i$ a basis of $(\R^4)^* \supset \tf^*$ dual to a basis of an $\R^4$ in which $\tf$ is embedded as the subspace annihilated by $\sum_i \be_i$.  Thus denoting the $A_3$ masses as $\bmu\in\C^4 \supset \tf_\C$, we have
\begin{align}\label{A3srbasis}
\ba_1(\bmu) &= \m_1-\m_2, &
\ba_2(\bmu) &= \m_2-\m_3, &
\ba_3(\bmu) &= \m_3-\m_4. 
\end{align}
The $\Z_2$ outer automorphism of $A_3$ is generated by an element, $o$, which exchanges $\a_1 \leftrightarrow \a_3$ and leaves $\a_2$ invariant.  This therefore acts on the $A_3$ masses as 
\begin{align}\label{A3outer-A3}
o &: (\m_1,\m_2,\m_3,\m_4)\mapsto -(\m_4,\m_3,\m_2,\m_1) .
\end{align} 
In terms of the $BC_3$ masses, this is the $\Z_2$ generated by 
\begin{align}\label{A3outer-BC3}
o &: (m_1,m_2,m_3) \mapsto (-m_1,m_2,m_3) ,
\end{align}
as follows from \eqref{BC3toA3}.

\subsection{The $\ff=\U(1)$ $\cN=3$ theories with $\D(u) = 6 ,4, 3$
\label{sec3.2}}

In \cite{Garcia-Etxebarria:2015wns}, Garc\'ia-Etxebarria and Regalado propose a novel F-theory construction of 4d field theories preserving $\cN=3$ supersymmetry.  These are necessarily isolated superconformal field theories with $U(3)$ R-symmetry group, Coulomb branch vevs of dimension $\Delta(u)=\ell$, $\ell\in\{3,4,6\}$, and central charges $a=c=(2\ell-1)/4$ \cite{Aharony:2015oyb,Nishinaka:2016hbw}. Furthermore, they also have no $\cN=3$ relevant deformations, and so, in particular, no continuous flavor symmetry.

The authors of \cite{Garcia-Etxebarria:2015wns} propose 3 series of such $\cN=3$ theories, labelled by $\ell\in\{3,4,6\}$ and $L\in\Z^+$.  These theories have moduli spaces $\cM_{\ell,L}=(\C^3/\Z_\ell)^L/S_L$, describing vevs of operators with scaling dimensions $n\ell$, $n\in\{1,\ldots,L\}$. In fact, \cite{Garcia-Etxebarria:2015wns} discusses evidence for multiple inequivalent versions of each of these theories, with the same low energy description, but presumably differing by some non-perturbative analog of discrete gauge factors \cite{Nishinaka:2016hbw}.  Here the ``rank", $L$, of these theories (the number of D3 brane probes of the F-theory geometry) controls the dimension of the moduli space.  We will concentrate on the $L=1$ case, since those will have rank-1 Coulomb branches when reinterpreted as $\cN=2$ theories.  We will call these the $\ell=3,4,6$ rank-1 $\cN=3$ theories.

Now let us reinterpret these theories as $\cN=2$ SCFTs.  The $U(3)$ $\cN=3$ R-symmetry decomposes as $U(1)_F\times U(2)_R$ with respect to a choice of $\cN=2$ subalgebra.  Here $U(1)_F$ is interpreted as an $\cN=2$ flavor symmetry and $U(2)_R$ is the $\cN=2$ R-symmetry.  Likewise, for $L=1$, the moduli space decomposes as $\cM_{\ell,1} = \cH_0 \cup (\cH_1 \times \C^*)$ where $\C^*$ is a 1-complex-dimensional Coulomb branch (minus the origin), $\cH_0$ a 1-quaternionic-dimensional Higgs branch over the origin, and $\cH_1$ is a 1-quaternionic-dimensional hyperk\"ahler fiber of a mixed branch over the generic points of the Coulomb branch.  Thus, as $\cN=2$ theories, the $\cN=3$ $\ell=3,4,6$ theories have 1-dimensional Coulomb branches with parameter $u$ of dimension $\D(u) = \ell$, and all have a $U(1)_F$ flavor symmetry, and thus a single mass deformation.  (Since $\D(u)>2$ there are no relevant or marginal chiral deformations.)  

In the table 1 of \cite{Argyres:2015gha} there is only one deformation of the $III^*$, and $IV^*$ singularities with a single mass parameter (i.e., rank-1 flavor algebra) while there are two possibilities in the $II^*$ case. The compatible identifications are: 
\begin{align}\label{ident}
\ell & = 6: & 
 &
\begin{cases}
II^*\to \{I_1, III^*\},\\
II^*\to \{I_2, IV^*_{Q=\ssqt}\},\\ 
\end{cases}& \ff_6 &= A_1 , 
\nonumber\\
\ell & = 4: & 
&\ \, III^* \to \{I_1, IV^*_{Q=1}\}, & \ff_4 &= A_1 ,
\\
\ell & = 3: & 
&\ \ \, IV^* \to \{I_1, I_1^*\}, & \ff_3 &= \U(1) .
\nonumber
\end{align}
The $\ell=3$ geometry's flavor symmetry matches the $\U(1)$ predicted for the $\cN=3$ theory,
\begin{align}\label{N3ell3flavor}
\ff_3 = \U(1).
\end{align}
As the first entry in \eqref{misident2} shows, the $\ell=6$ and $\ell=4$ geometries can instead be reinterpreted as having flavor symmetry
\begin{align}\label{N3ell64flavor}
\ff_6 = \ff_4 = \Z_2\ltimes\U(1),
\end{align}
where the $\Z_2$ is the outer automorphism which acts on $\U(1)$ by reversing the sign of charges, and so as $m\mapsto -m$ on the complex mass.  So, if this identification with the $\cN=3$ SCFTs is correct, we predict that the flavor groups of the $\ell=6$ and $4$ theories have the extra discrete $\Z_2$ factor shown in \eqref{N3ell64flavor}. This holds true for both identifications of the $\ell=6$ $\cN=3$ theory presented in \ref{ident} which also have the same values for the $a$ and $c$ central charges (see \ref{tab1}).  The only way to distinguish these two inequivalent identifications from their low energy data is through their RG flows (as explained in more detail in section \ref{sec3.3.2}). 

This discrete $\Z_2$ flavor factor can in fact be seen in the F-theory construction of the SCFTs in \cite{Garcia-Etxebarria:2015wns}.  In that construction, the moduli space of the rank-1 theories is $\cM_{\ell,1} = \C^3/\Z_\ell$, given by $(z_1,z_2,z_3)\in\C^3$ with 
\begin{align}\label{}
\Z_\ell :& (z_1,z_2,z_3) \mapsto (e^{+2\pi i /\ell} z_1, e^{+2\pi i /\ell} z_2, e^{-2\pi i /\ell} z_3).
\end{align}
$(z_1,z_2,\bar z_3)$ transform as a triplet under the $\cN=3$ $U(3)$ R-symmetry.  Take the $z_2=z_3=0$ subspace to be the $\cN=2$ CB, so that $u:=z_1^\ell$ is its $\Z_\ell$-invariant coordinate.  Then $(z_2,\bar z_3)$ transform as a doublet under the $\cN=2$ $SU(2)_R$ symmetry, and the Higgs branch is $\C^2/\Z_\ell$ with $\C^2$ the $z_{2,3}$-plane.  In terms of $\Z_\ell$-invariant holomorphic coordinates it is given by
\begin{align}\label{N3Higgs}
\cH_0 &\simeq \C^3/\vev{VW-X^\ell},&
\text{with}\quad V &:=z_2^\ell,
\quad  W:=z_3^\ell,
\quad \text{and}\quad  X:=z_2 z_3.
\end{align}
Similarly, the fiber of the mixed branch over a generic point (e.g., $z_1^\ell=1$) of the CB is simply a copy of the $z_{2,3}$-plane $\simeq \C^2$.  Since the Higgs branch operators are neutral under $U(1)_R$, it follows that the $U(1)_R$ is the subgroup of the $U(3)$ which leaves $z_2$ and $z_3$ invariant, so acts by phase rotations of $z_1$ only.

The $U(1)_F$ does not act on the CB, and acts as a tri-holomorphic isometry of the Higgs branch hyperk\"ahler structure.  This means that it acts holomorphically on $(z_2,z_3)$ and preserves the K\"ahler and (2,0) forms on $\cH_0$.  These are the ones inherited from the complex structure and flat metric on $\C^3$.  This implies the K\"ahler form is $\w^{(1,1)} \propto dz_2 \wedge d\bar z_2 + dz_3 \wedge d\bar z_3 = |V|^{2(1-\ell)/\ell}dV \wedge d\bar V + |W|^{2(1-\ell)/\ell}dW \wedge d\bar W$, and the holomorphic 2-form is $\w^{(2,0)} \propto dz_2 \wedge dz_3 = X^{1-\ell} dV \wedge dW$.  Thus for $U(1)_F$ to be a tri-holomorphic isometry, it must be the $U(1)$ action for which the coordinates have charges $F(z_1)=0$, $F(z_2)=-F(z_3)=1$ (in an arbitrary normalization).  Thus 
\begin{align}\label{U1F}
U(1)_F: \qquad F(V)=-F(W)=\ell,
\quad\text{and}\quad F(X)=0.
\end{align}

There are also tri-holomorphic isometries disconnected from the identity which exchange $V$ and $W$.\footnote{We thank Y. Tachikawa for pointing this out to us.}  It is not hard to see that these can only be of the form $(V,W,X) \mapsto (e^{i\a} W, e^{i\b} V, e^{i\g} X)$, for phases satisfying $\a+\b=\g \ell$ from \eqref{N3Higgs} and $\a+\b+\g(1-\ell)=\pi$ from preserving $\w^{(2,0)}$.  This implies $\g=\pi$ and we can use the $U(1)_F$ action \eqref{U1F} to rotate phases to $\a=0$ and $\b=\pi \ell$, thus giving a discrete isometry
\begin{align}\label{ptr}
p: (V,W,X) \mapsto (W, (-)^\ell V,  - X).
\end{align}
For $\ell$ even, $p$ generates a $\Z_2$, while for $\ell$ odd it generates a $\Z_4$.  

For $\ell$ even, since $p$ interchanges $V$ and $W$ which have opposite $U(1)_F$ charges, it is plausible that the $\Z_2$ acts as the outer automorphism on $U(1)_F$ flipping the sign of the linear mass parameter $m\to -m$. The latter action also realizes the sign flip on the $X$ as it can be seen from the form of the mass term $\sim m X$ which appears in the lagrangian.\footnote{Recall that the full $\cN=2$ mass term has the form $m (\tilde Q^2)_{I=1} \hat{\cal B}_1 + c.c.$ \cite{Argyres:2015ffa}. The $X$ operator should be identified with the momentum map which is the highest weight component of the $SU(2)_R$ triplet in the $\hat{\cal B}_1$ multiplet.}

The situation is different for $\ell$ odd.  The discrete isometry $p$ in \eqref{ptr} cannot be interpreted as a flavor symmetry since $p^2=-I$ on $(V,W)$ which is the action of the center of $SU(2)_R$.  So the $p$ action on $X$ must instead be interpreted as a composition of an $SU(2)_R$ transformation, $r:\ X\to-\bar{X}$, with CPT conjugation, $c:\ X\to \bar{X}$.  This realizes $X\to-X$ as a non-flavor symmetry action.  As far as $X$ goes the two actions are indistinguishable, yet they have different actions on $(W,V)$.  In particular:
\begin{align}\label{}
r: \begin{pmatrix} z_2\\ \bar z_3\end{pmatrix}
&\mapsto \begin{pmatrix} 0&1\\-1 & 0\end{pmatrix}
\begin{pmatrix} z_2\\ \bar z_3\end{pmatrix},&
\Rightarrow \qquad\qquad
r: (V,W,X) &\mapsto (\bar W, -\bar V, -\bar X) ,
\nonumber\\
c: \begin{pmatrix} z_2\\ z_3\end{pmatrix}
&\mapsto \begin{pmatrix} \bar z_2\\ \bar z_3\end{pmatrix},&
\Rightarrow \qquad\qquad
c: (V,W,X) &\mapsto (\bar V, \bar W, \bar X) .
\end{align}
The transformations above follow since $V$ and $\bar W$ transform as the highest- and lowest-weight components of a spin-$\ell/2$ representation of $SU(2)_R$, while $X$ is the highest-weight component of a spin-1 representation (i.e., the moment map for the Higgs branch).  The composition of $r$ and $c$ reproduces \eqref{ptr}. Thus for $\ell$ odd we are led to the following identification:
\begin{align}\label{}
p &= c \circ r,& \text{with} \quad
c&\in \Z_2 \ \text{charge conjugation},& \text{and}\quad 
r&\in \Z_4 \subset SU(2)_R.
\end{align}    
Since $r$ is in $SU(2)_R$ it does not act on the mass parameters.  It follows that $p$ acts on the mass parameters as $m\mapsto \bar m$ from the charge conjugation, and thus should not be identified as the $U(1)_F$ outer automorphism $m\mapsto -m$. This non-holomorphic action on the masses is not visible in the CB geometry, and so for $\ell=3$ we expect the flavor symmetry visible in the deformed CB geometry is simply $\ff_3 = \U(1)$, as in \eqref{N3ell3flavor}.

\section{Some checks}\label{sec3.3}

There is further evidence supporting the identifications discussed above.  This evidence comes from: comparing the discriminant locus of the curve in \cite{Chacaltana:2016shw} with ours;  revisiting the RG flow consistency conditions (for details see \cite{Argyres:2015ffa,Argyres:2015gha}) in light of the new flavor groups; and central charge computations using the technique of \cite{Shapere:2008zf} as described in \cite{Argyres:2015ccharges}. 

\subsection{Curve discriminants
\label{sec3.3.1}}

Let us start by comparing the curve in \cite{Chacaltana:2016shw} with the one we constructed in \cite{Argyres:2015gha} but with the newly identified flavor group.  In particular we can explicitly check that once we turn on the same mass deformations the locations of the singularities for the two curves coincide. 

When the $A_1\oplus A_1 \subset A_3$ mass deformations are turned on, the curve for the $[II^*,A_3\rtimes \Z_2]$ CB is given in \cite{Chacaltana:2016shw} by
\begin{align}\label{CDTcurve}
F(x,u,z):=
\l^8 +\f_2 \l^6 + \f_4 \l^4 + \f_6 \l^2 + {\til\f}^2
=0,
\end{align}
where
\begin{align}\label{subA3}
\l &= x\, dz,&
\f_2 &=\frac{ 
M_+^2 (z - z_1) z_{12} z_{23} + M_-^2 (z - z_2) z_{21} z_{13}}{(z - z_1)^2 (z - z_2)^2 (z - z_3)}\, dz^2 ,
\\
\f_4 &= \frac14 {\f_2}^2,&
\f_6 &= u \frac{z_{12}^4 z_{13} z_{23}}
{(z -z_1)^5 (z - z_2)^5 (z - z_3)^2}\, dz^6, 
\qquad \qquad
\til\f =0.
\nonumber
\end{align}
Here $z$ is a coordinate on the Riemann sphere, $x$ is a coordinate on the cotangent space to the sphere, $\{z_1,z_2,z_3\}$ are the (arbitrary) locations of the three punctures, $z_{ij} := z_i - z_j$, $u$ is the CB coordinate, and $M_\pm$ are the linear masses of the two $A_1$ flavor factors.  (In particular, in terms of the $A_1$ Casimirs, $m_2$ and $m_2'$, introduced in \cite{Chacaltana:2016shw}, $M_+^2 = m_2$ and $M_-^2 = -m_2'$.)

This curve \eqref{CDTcurve} is singular for values of $u$ solving $F=\del F/\del x=\del F/\del z=0$.  Locating the punctures at $z_1=1$, $z_2=-1$, and $z_3=0$ for convenience, we find that the curve has three singular loci in the $u$ plane located at the zeros of the polynomial\footnote{This is not quite the discriminant of the curve since we did not determine the multiplicity of its zeros.}
\begin{align}
D &= u \Bigl(
u^2 
+\frac{1}{27} 
(M_+^2-2 M_-^2)(2 M_+^2-M_-^2) (M_+^2+M_-^2) u
-\frac{1}{108}
M_+^4 M_-^4 (M_+^2-M_-^2)^2 
\Bigr).
\end{align}

We now repeat this analysis for the curve found in \cite{Argyres:2015gha}.  This is a straightforward procedure once we take care of two subtleties.  First, the curve reported in the appendix of \cite{Argyres:2015gha} is written in terms of the linear masses $(m_1,m_2,m_3)$ associated with the wrong ($BC_3$) flavor group.  The $A_3$ form of the curve is given by using \eqref{BC3toA3} to rewrite the curve in terms of $A_3$ linear masses $(\m_1,\m_2,\m_3,\m_4)$ satisfying $\sum_i \m_i=0$.  

The second subtlety is to identify the directions in the $(\m_1,\m_2,\m_3,\m_4)$ mass deformation space corresponding to turning on only the $M_\pm$ mass parameters in \eqref{CDTcurve}.   The manifest $A_1\oplus A_1 \subset A_3$ flavor symmetry of the curve \eqref{CDTcurve} has two quadratic mass Casimirs, $M_\pm^2$, while the full $A_3$ symmetry has three independent Casimirs which we can take to be $N_a = \sum_{i=1}^4 \mu_i^a$, for $a=2,3,4$.  So if only $A_1\oplus A_1$ masses are turned on, only $N_2$ and $N_4$ can be non-zero, and we must have
\begin{align}\label{}
N_3  = -3(\m_1+\m_2)(\m_1+\m_3)(\m_2+\m_3) = 0.
\end{align}
Take the solution
\begin{align}\label{A3toA1A1break}
\m_1 = -\m_2
\end{align}
so that, with respect to the basis of simple roots of $A_3$ in \eqref{A3srbasis}, the $A_3$ outer automorphism  \eqref{A3outer-A3} acts non-trivially on the chosen $A_1\oplus A_1$ subgroup.\footnote{The choice $\m_1=-\m_3$ would have worked equally well.  The choice $\m_2=-\m_3$ would have required a different choice of outer automorphism action.  Recall that outer automorphisms are only defined up to the action of inner automorphisms, which can be thought of as changing the choice of basis of simple roots that the outer automorphism acts on.}

Then, writing the SW curve of appendix A.1.4 of \cite{Argyres:2015gha} in terms of the $\m_i$ using \eqref{BC3toA3} and substituting for $\m_1$ using \eqref{A3toA1A1break}, the  resulting curve becomes singular at the zeros of the discriminant
\begin{align}\label{Dpdisc}
D' &= u^8 \Bigl(
u^2 
+\frac{1}{2} (\m_2^2-2 \m_3^2)(2 \m_2^2-\m_3^2)
(\m_2^2+\m_3^2) u
-\frac{27}{16} \m_2^4 \m_3^4(\m_2^2-\m_3^2)^2
\Bigr).
\end{align}
The discriminants, $D$ and $D'$, of the two curves clearly agree after identifying their linear mass parameters as
$M_+ = (\sqrt[6]{2}/\sqrt{3}) \, \m_2$ and $M_- = (\sqrt[6]{2}/\sqrt{3}) \, \m_3$.

As mentioned above the $\Z_2$ factor of the flavor symmetry should be identified with the interchange of the two $A_1$ factors. From the explicit expression of $M_\pm$ in terms of the $\mu_{2,3}$ we see that this action is in fact compatible with the action of the outer automorphism of the full $A_3$ as identified in \eqref{A3outer-A3}.

We cannot perform a similar discriminant check for the $\ell=3,4,6$ $\cN=3$ theories because it is not clear how to modify the string construction in \cite{Garcia-Etxebarria:2015wns} to turn on the $\cN=2$ $\U(1)$ mass deformation.

\subsection{RG flows
\label{sec3.3.2}}

In \cite{Argyres:2015gha} we claimed that the $[II^*,BC_3]$ theory did not pass the RG flow condition if the frozen $I_1^*$ singularity was interpreted as a lagrangian field theory.  The RG flow test depends on the identification of the global flavor group. Thus we should redo the analysis of minimal adjoint flavor breaking RG flows for the $[II^*,A_3\rtimes\Z_2]$ theory.  $A_3$ has two inequivalent minimal adjoint breakings, one from turning on a vev for either node at the end of the Dynkin diagram, and one for turning on a vev for the middle node.  Keeping track of the discrete $\Z_2$ factor as well, it is easy to see that these give rise to the following flavor breakings,
\begin{align}\label{A3Z2mab}
A_3\rtimes \Z_2 \ 
\begin{cases}
\to A_2 \oplus \U(1)                
&\to \{I^*_1,I_3\}, \ \cmark\\
\to \U(1) \oplus (A_1\oplus A_1)\rtimes\Z_2  
&\to \{I_3^*,I_1\}. \ \cmark
\end{cases}
\end{align}
For each flavor breaking we have also recorded deformation pattern of the parent $II^*$ singularity which results from putting in the specific breaking masses in the $[II^*, A_3 \rtimes \Z_2]$ SW curve described in section \ref{sec3.1}.

In the first line of \eqref{A3Z2mab} the $I_1^*$ is frozen while the $I_3$ must be interpreted as a $\U(1)$ theory with three charge one hypermultiplets providing a $\U(3)\equiv A_2\oplus\U(1)$ flavor symmetry.  In the second line, the $I_1$ provides a $\U(1)$ flavor factor while the $I_3^*$ should be interpreted as an $\SU(2)$ w/ $4\cdot{\bf2}+1\cdot{\bf4}$ lagrangian theory with $\SO(4)\simeq A_1\oplus A_1$ flavor symmetry and charge normalization $a=1$.  (For details on these identifications see \cite{Argyres:2015ffa,Argyres:2015gha}.)  Since these IR singularities precisely reproduce the expected unbroken flavor symmetries, we conclude that the $[II^*,A_3\rtimes\Z_2]$ theory passes the RG flow consistency condition.

Once the existence of the $[II^*,A_3\rtimes\Z_2]$ theory is accepted, any other SCFTs it flows to must also be consistent.  We will now check that that is the case.  

In \cite{Argyres:2015gha} we found that one of the $[II^*,BC_3]$ minimal adjoint breakings generates the deformation $II^*\to\{III^*,I_1\}$.  This direction is no longer a minimal adjoint breaking in the $A_3\rtimes\Z_2$ interpretation of the theory but instead corresponds to setting $\m_1=\m_2=0$ and $\m_3=-\m_4$ in the $A_3$ linear masses defined in \eqref{BC3toA3}.  Along this direction we expect an unbroken $A_1\oplus\U(1)^2$ flavor group.  Because the $I_1$ only contributes a $\U(1)$ factor, the remaining part should be identified as the flavor group of the CFT at the $III^*$ singularity.  

In table 1 of \cite{Argyres:2015gha} the only deformation of a $III^*$ singularity with a rank 2 flavor group is the $[III^*,A_1\oplus A_1]$ curve (which also failed the RG flow test for a lagrangian interpretation of the $I_1^*$ singularity).  But our table \eqref{misident} of possible flavor misidentifications allows for this curve to be interpreted instead as the curve of a $[III^*,A_1\oplus(\U(1)\rtimes\Z_2)]$ theory.  Notice that the $\Z_2$ factor of the initial $II^*$ singularity is broken along this RG flow direction, thus the $\Z_2$ factor of the $III^*$ is a new one.

We should now redo the RG flow analysis for the newly identified flavor group:
\begin{align}\label{A1U1Z2mab}
A_1\oplus (\U(1)\rtimes\Z_2) \ 
\begin{cases}
\to A_1\oplus \U(1)  &\to \{I_1^*,I_2\}, \ \cmark\\
\to \U(1)\oplus (\U(1)\rtimes\Z_2)  &\to \{I^*_2,I_1\}. \ \cmark
\end{cases}
\end{align}
In the first case the $I_1^*$ is frozen while the $I_2$ provides the non-abelian, $A_1$, component of the flavor group. The second case now also passes the RG flow test since the $I_1$ provides one $\U(1)$ factor while the $I_2^*$ must be interpreted as an $\SU(2)$ w/ $2\cdot {\bf2}+1\cdot {\bf4}$ gauge theory with $\SO(2)\simeq\U(1)$ flavor group with charge normalization $a=1$.  We thus conclude that the $[III^*,A_1\oplus(\U(1)\rtimes\Z_2)]$ passes the RG flow condition as well.

Next, we can study the flow from $[III^*, A_1\oplus (\U(1) \rtimes \Z_2)]$ to the $[IV^*,\U(1)]$ theory, which, as we argued earlier, can be identified with the $\ell=3$ $\cN=3$ theory.  The right RG flow direction was already identified in \cite{Argyres:2015gha} when we studied the non-adjoint breaking
\begin{align}\label{IIIsA1A1-RG}
[III^*, A_1\oplus (\U(1)\rtimes\Z_2)]
\to \{ IV^*, I_1  \} \quad \text{for}\quad 
m_1= (i/\sqrt{3})\,m_2.
\end{align}
The $\Z_2$ part of the flavor group of the $III^*$ SCFT acts by flipping the sign of $m_2$ and it is thus broken along the flow to the $[IV^*,\U(1)]$ theory, providing a beautifully consistent picture.


It would be interesting to know if there is a different class-$\cS$ construction of the $[II^*,A_3\rtimes\Z_2]$ theory for which all three mass deformation parameters are realized.  If there were, it would imply that the two other $I_1^*$-series rank-1 SCFTs --- the $[III^*,A_1\oplus(\U(1)\rtimes \Z_2)]$ and the $\cN=3$ $[IV^*,\U(1)]$ theories --- are accessible via class-$\cS$ constructions.

We can analyze in a similar manner the RG flows involving the other two ($\ell=4,6$) $\cN=3$ theories.  As discussed above, there are two theories which are compatible with the properties of $II^*$ $\cN=3$ theories \eqref{ident}.  Both are $[II^*,\U(1)\rtimes\Z_2]$ theories with a single mass parameter.  Turning such mass deformation on splits the $II^*$ singularity as $II^*\to\{I_1,III^*\}$ and $II^*\to\{I_2,IV^*_{\ssqt}\}$ respectively.  It was argued in \cite{Argyres:2015ffa, Argyres:2015gha} that both the $III^*$ and the $IV^*_{Q=\ssqt}$ singularity on the $[II^*,\U(1)\rtimes\Z_2]$ CB must be identified with new frozen SCFTs: $[III^*,\varnothing]$ and $[IV^*,\varnothing]_{Q=\ssqt}$.  These frozen theories are $\cN=2$ SCFTs with rank-1 CBs and with no relevant $\cN=2$ deformations (and hence empty flavor symmetry: $\ff=\varnothing$). The dimensions of their CB parameters are $\D(u)=3$ and $4$, respectively.  Furthermore the Dirac quantization condition \cite{Argyres:2015ffa} implies that the BPS spectrum of the $IV^*_{Q=\ssqt}$ should consist of states with electric and magnetic charges proportional to $Q=\sqrt{2}$.  There are no further RG flow consistency checks that can be performed on these theories, but, as we will see in sections \ref{sec3.3.3} and \ref{sec3.3.4}, we can use these flows and knowledge of the Higgs branches and central charge relations of the $\ell=6$ $\cN=3$ theory to constrain the central charges and Higgs branches of the new frozen SCFTs $[III^*,\varnothing]$ and $[IV^*,\varnothing]_{Q=\ssqt}$.

Similarly, the $\ell=4$ or $[III^*,\U(1)\rtimes\Z_2]$  theory has a single mass parameter which deforms the $III^*$ singularity as $III^* \to \{ I_1, IV^*\}$.  It was argued in \cite{Argyres:2015ffa, Argyres:2015gha} that this $IV^*$ singularity on the $[III^*,\U(1)\rtimes\Z_2]$ CB must be identified with a new frozen rank-1 $[IV^*, \varnothing ]_{Q=1}$ SCFT with dimension $\D(u)=3$ CB parameter.  As in the $\ell=6$ case, there are no further flows from the $[III^*,\U(1)\rtimes\Z_2]$ to check.  But now there is a possible SCFT with a $II^*$ singularity and a rank-2 flavor group which can flow to the $[III^*,\U(1)\rtimes\Z_2]$  theory.   This $II^*$ theory has generic deformation pattern $II^*\to\{{I_1}^2,IV^*\}$ with the $IV^*$ singularity identified with the frozen $[IV^*,\varnothing]_{Q=1}$ SCFT \cite{Argyres:2015ffa}.  The flavor group of this new $II^*$ SCFT was identified in \cite{Argyres:2015gha} as $\ff=G_2$.  However, according to the discussion in section \ref{sec2}, a $G_2$ flavor symmetry could instead be interpreted according to \eqref{misident} and \eqref{weylinfo} as the smaller symmetries $A_2 \rtimes \Z_2$ or $\U(1)^2 \rtimes (\Z_2 \ltimes S_3)$.  Thus there are three candidate SCFTs, $[II^*,G_2]$, $[II^*,A_2\rtimes\Z_2]$, and $[II^*,\U(1)^2 \rtimes (\Z_2 \ltimes S_3)]$, which could flow to the $\ell=4$ $\cN=3$ $[III^*,\U(1)\rtimes\Z_2]$ theory.  As discussed in section \ref{sec4} and shown in figure \ref{IVs}, the RG flow from the $[II^*,G_2]$ theory is not consistent, while the flows from the other two theories are.  So we are not able to determine which of $[II^*, A_2\rtimes\Z_2]$ or $[II^*,\U(1)^2 \rtimes (\Z_2 \ltimes S_3)]$ are ``RG parents" of the $[III^*,\U(1)\rtimes\Z_2]$ theory.   In section \ref{sec4}, however, we will see by examining the full space of RG flows that the $[II^*,A_2\rtimes\Z_2]$ is a ``good" theory while the $[II^*,\U(1)^2 \rtimes (\Z_2 \ltimes S_3)]$ theory is ``ugly" (it requires accidental flavor symmetry enhancements in the IR).  For this reason we show only the $[II^*,A_2\rtimes\Z_2]$ theory in table \ref{tab1}, and will discuss only its central charges and Higgs branches below.  

We emphasize that the $[II^*, \U(1)^2 \rtimes (\Z_2 \ltimes S_3)]$ theory is \emph{not} logically excluded: we have excluded it only to keep our discussion relatively short.
We will discuss consistency of these flows as well as of flows among rank-1 SCFTs with all possible flavor symmetry assignments in section \ref{sec4}.

\subsection{Central charges
\label{sec3.3.3}}

We will summarize here how the $a$, $c$, and $k$ central charges of rank-1 $\cN=2$ SCFTs can be computed from a generalization of the argument developed by Shapere and Tachikawa in \cite{Shapere:2008zf}.  The $a$ and $c$ central charges of the 4d conformal algebra are certain coefficients in OPEs of energy-momentum tensors, and the $k$ central charges appear in the OPEs of flavor currents.  We use the standard normalizations of the central charges where for $n_v$ free vector multiplets and $n_h$ free hypermultiplets transforming under a nonabelian global symmetry $\ff$,
$24 a = 5 n_v + n_h$, $12 c = 2 n_v + n_h$, and $k = T({\bf 2n_h})$.  Here $\bf 2n_h$ is the representation of $\ff$ under which the half-hypermultiplets transform.  The quadratic index is defined as $T(\br) := [\text{rank}(\ff)]^{-1} \sum_{\l\in\br} (\l,\l)$, where the weights are normalized so that the long roots of $\ff$ have length-squared 2.  In this normalization $T(\bn)=1$ for $\SU(n)$.

We obtain the $a$, $c$, and $k$ central charges of each entry of table \ref{tab1} as a function of a few parameters involving mostly data from the deformation pattern singularities.  These can be used both as checks for the correctness of the identifications made in sections \ref{sec3.1} and \ref{sec3.2}, and also to deduce more information about the various SCFTs in table \ref{tab1} and their RG flows.

\subsubsection*{The \textit{a} and \textit{c} central charges}

The following formulas for $a$ and $c$ are derived in  \cite{Argyres:2015ccharges}:
\begin{align}\label{CCs}
24 a &=5+h_1+6(\D-1)+\D\sum_{i=1}^Z N_i,\\
12 c &=2+h_1+\D\sum_{i=1}^Z N_i.\nonumber
\end{align}
Here $\D=\D(u)$ is the scaling dimension of the CB parameter, and $h_1$ is the quaternionic dimension of the Higgs fiber of the ``enhanced Coulomb branch" (ECB) of the SCFT.  $Z$ and $N_i$ refer to properties of the generic mass deformation of the SCFT.  In particular, $Z$ counts the number of undeformable Kodaira singularities the initial singularity of the SCFT splits into upon turning on a generic relevant deformation, and $N_i$ is the central charge contribution of the conformal or IR-free theory corrsponding to the $i$th such singularity.  It is given by \cite{Argyres:2015ccharges}
\begin{align}\label{Nis}
N_i:=\frac{12c_i-h_i-2}{\D_i} ,
\end{align}
where $c_i$, $h_i$ and $\D_i$ are respectively the $c$ central charge, the quaternionic ECB dimension, and the CB scaling dimension of the SCFT or IR free field theory corresponding to the $i$-th Kodaira singularity in the deformation pattern.  When these undeformable singularities have a lagrangian interpretation, $N_i$ is easily computable.  For undeformable $I_n$ singularities $N_{I_n}=1$ while for a frozen $I_1^*$ singularity, $N_{I_1^*}=3$; see \cite{Argyres:2015ccharges} for the details.

Since ECBs might not be familiar, we pause to summarize their main properties;  the structure of ECBs is discussed in more detail in \cite{Argyres:2015ccharges}.  ``Enhanced Coulomb branch" is our name for a mixed Higgs-Coulomb branch that occurs over the whole CB; thus the CB proper is a sub-variety of the ECB, and the ECB is in effect an enlarged Coulomb branch.  The ECB locally has a direct product geometry $U_i \times \cH_1$ where $\{U_i\}$ is an open covering of the regular points of the CB, and $\cH_1$ is a hyperk\"ahler space.  $h_1$ is the quaternionic dimension of $\cH_1$, so the total complex dimension of the ECB is $2h_1+1$ (since we are discussing here only theories with rank-1 CBs).  Over a generic point on the CB, the $2h_1$ complex scalars whose vevs parameterize the ECB fiber are neutral under the low energy electromagnetic $\U(1)$ gauge group, so the ECB fiber over a regular CB point is a flat hyperk\"ahler space, $\cH_1 = \H^{h_1}$.  The moduli spaces of $\cN=4$ theories as well as of the $\cN=3$ SCFTs described in section \ref{sec3.2} are examples of ECBs.  But ECBs commonly occur in $\cN=2$ field theories as well.  Even when there is an ECB, there can be additional mixed and Higgs branches.  In the case of a SCFT with rank-1 CB, the only possibility for an additional branch is a Higgs branch, $\cH_0$, which is a hyperk\"ahler cone with tip touching the CB at its singular point (the ``origin").  It is a logical possibility that $\cH_0$ might have multiple components and that the intersection of $\cH_0$ with the $\cH_1$ fiber of the ECB over the origin might be any hyperk\"ahler cone from the empty one (the origin istelf) to all of $\cH_1$.

We will now apply \eqref{CCs} and \eqref{Nis} to the $[II^*,A_3\rtimes\Z_2]$, $[II^*,\U(1)\rtimes\Z_2]$, $[III^*,\U(1)\rtimes\Z_2]$, and $[IV^*,\U(1)]$ SCFTs discussed above.  

\vspace{2mm}
\begin{flalign*}
[II^*,A_3\rtimes\Z_2]: &&
a&= \frac{25}{8}, & 
c&= \frac{7}{2}, &
h_1 &= 4 .&&\qquad
\end{flalign*}
The $a$ and $c$ central charges for this theory were  computed in \cite{Chacaltana:2016shw} to be $24 a=75$ and $12c=42$.  Plugging into \eqref{CCs} using its deformation pattern $II^*\to\{{I_1}^3,I_1^*\}$ and that $\D(u)=6$, one finds that $h_1=4$.  Thus we make a prediction that the $[II^*,\SU(4)\rtimes\Z_2]$ has a 4 quaternionic dimensional ECB fiber.  It is worth noting that the fact that $h_1$ comes out as an integer is a non-trivial check of the corectness of our identification.  A sharper check will be found when we compute the flavor central charge, $k$, below.  It would also be interesting to determine the value for $h_1$ independently from the superconformal index of this theory, or by embedding this theory in a web of S-dualities.

\vspace{2mm}
\begin{flalign*}
[II^*,\U(1)\rtimes\Z_2]_{\{III^*,\ IV^*_{Q=\ssqt}\}}: &&
a&= \frac{11}{4}, & 
c&= \frac{11}{4}, &
h_1 &= 1 .&&\qquad
\end{flalign*}
We proposed that these curves are identified with the $\ell=6$ $\cN=3$ theory.  As explained in \cite{Aharony:2015oyb}, $\cN=3$ supersymmetry requires $a=c$.  This, together with $\D(u)=6$ and \eqref{CCs} determine $a = c = 11/4$.  Furthermore, as reviewed in section \ref{sec3.2}, it also implies that this theory has a one-quaternionic-dimensional ECB fiber, thus $h_1=1$.  The deformations patterns of these theories are $II^*\to\{III^*,I_1\}$ and $II^*\to\{IV^*_{Q=\ssqt},I_2\}$ where the $III^*$ and $IV^*_{Q=\ssqt}$ singularities must be identified with rank-1 isolated SCFTs $[III^*,\varnothing]$ and $[IV^*,\varnothing]_{Q=\ssqt}$. Thus  \eqref{CCs} determine a relation between $c_{III^*}/c_{IV_{Q=\ssqt}^*}$ and $h_{III^*}/h_{IV_{Q=\ssqt}^*}$ of these non-lagrangian theories from which it follows:
\begin{flalign*}
[III^*,\varnothing]:& &
a &=\frac{13}{8},&
c &=\frac{3}{2}, &
h_1 &=  0,&&\qquad
\end{flalign*} 
\begin{flalign*}
[IV^*,\varnothing]_{Q=\ssqt}: &&
a &=\frac{29}{24},&
c &=\frac{7}{6}, &
h_1 &=  0,&&\qquad
\end{flalign*}
where we have assumed (as will be justified in section \ref{sec3.3.4}) that for both theories above there is no ECB, i.e., $h_1=0$. 

\vspace{2mm}
\begin{flalign*}
[III^*,\U(1)\rtimes\Z_2]: &&
a&= \frac{7}{4}, & 
c&= \frac{7}{4}, &
h_1 &= 1 .&&\qquad
\end{flalign*}
This theory is identified with the $\ell=4$ $\cN=3$ theory.  Just as in the previous case we get $a =c = 7/4$ and $h_1 = 1$, and 
\begin{flalign*}
[IV^*,\varnothing]_{Q=1}: &&
a &=\frac{55}{48},&
c &=\frac{25}{24}, &
h_1 &=  0,&&\qquad
\end{flalign*}
for the frozen non-lagrangian SCFT that it flows to.

\vspace{2mm}
\begin{flalign*}
[IV^*,\U(1)]: &&
a&= \frac{5}{4}, & 
c&= \frac{5}{4}, &
h_1 &= 1 .&&\qquad
\end{flalign*}
This theory is identified with the $\ell=4$ $\cN=3$ theory.  As in the previous two cases we get from $\cN=3$ supersymmetry the central charges and ECB fiber dimension shown above.  Unlike the previous two cases, however, this theory's deformation pattern, $III^*\to\{I_1,I_1^*\}$, is to IR free lagrangian theories, and so one can independently compute $c$ from \eqref{CCs} to obtain the same answer.  This is a strong indication that this theory should be identified as the $\ell=3$ $\cN=3$ theory constructed in \cite{Garcia-Etxebarria:2015wns}.

Finally, we note that the $a=c$ central charges of the $\cN=3$ theories found here agree with those found in \cite{Nishinaka:2016hbw}, who also find further evidence in support of those values coming from the structure of the chiral algebras associated to the Schur operators of those theories \cite{Beem:2013sza}.

\subsubsection*{Flavor central charges}

As explained in \cite{Shapere:2008zf}, the flavor central charges for $\U(1)$ factors of flavor groups are difficult to determine because of the possibility of them mixing under RG flows with the low energy global electric and magnetic $\U(1)$'s on the CB.  So we restrict ourselves to computing the flavor central charges, $k$, for nonabelian factors of the flavor symmetry.  Also, we can no longer use the strategy of turning on a generic mass deformation to compute $k$ since under such a deformation the low energy flavor group is entirely broken to $\U(1)$ factors.   Thus we must instead use special (e.g., minimal adjoint breaking) mass deformations which leave some nonabelian subgroup of the SCFT flavor symmetry unbroken.  

Let's say that under one such special mass deformation, $m$, our $[K,\ff]$ SCFT  (with $K$ the Kodaira type and $\ff$ the flavor symmetry) deforms to $Y$ singularities as
\begin{align}\label{}
[K,\ff] \xrightarrow{m} \left\{ [K_1, \ff_1], \ldots, [K_Y,\ff_Y] .
\right\}
\end{align}
Consequently the flavor symmetry breaks to $\ff \xrightarrow{m} \oplus_{i=1}^Y \ff_i$.  Ignoring any $\U(1)$ factors in this breaking, put the (topologically twisted) theory in a background of $n_i$ instantons for each (nonabelian) $\ff_i$.  This corresponds to a total $n$-instanton background for the original $\ff$ flavor symmetry where $n = \sum_{i=1}^Y n_i d_i$, and the $d_i$ are the Dynkin indices of embedding $\ff_i \hookrightarrow \ff$.  Then, as long as one knows the flavor central charges, $k_i$, for the $[K_i,\ff_i]$ theories (e.g., if they are lagrangian theories) one deduces from the arguments of \cite{Shapere:2008zf} that \cite{Argyres:2015ccharges}
\begin{align}\label{kform}
k &= \frac{\D}{\D_i} \left(\frac{k_i}{d_i} - T\left({\bf 2h^{(i)}_1}\right)\right) + T({\bf 2h_1}) ,&
&\text{for all $i$ such that $\ff_i$ is nonabelian.}
\end{align}
Here, as usual, $\D=\D(u)$ is the scaling dimension of the CB parameter.  $T({\bf 2h_1})$ is the quadratic index of the representation $\bf 2h_1$ of $\ff$ given by the representation of the flavor symmetry under which the $2h_1$ complex scalars of the hypermultiplets on the ECB fiber $\cH_1$ transform.  Similarly, $\D_i$ and $h_1^{(i)}$ are the CB scaling dimension and ECB fiber dimension of the $[K_i,\ff_i]$ theory and again $d_i$ are the Dynkin indices of embedding $\ff_i \hookrightarrow \ff$.  

We now apply \eqref{kform} to the $[II^*,A_3\rtimes\Z_2]$ theory.  The minimal adjoint breaking $A_3 \to A_2\oplus\U(1)$ mass deformation, $m_1$, deforms the singularity as
\begin{align}\label{}
[II^*,A_3\rtimes\Z_2] \xrightarrow{m_1} \left\{ [I_1^*, \varnothing]\, ,\,  [I_3,A_2\oplus\U(1)] \right\} .
\end{align}
Since the non-abelian flavor factor appearing in the second $(I_3)$ singularity is $A_2 \subset A_3$ with index of embedding $1$, we set $d_2=1$ in \eqref{kform}.  The $I_3$ singularity is an IR free $\U(1)$ gauge theory with 3 massless charge-1 hypermultiplets transforming in the $\bf 3$ of the $A_2$ flavor symmetry.  They thus contribute $k_2 = T({\bf 3}\oplus\bar{\bf 3}) = 2$ to the $A_2$ flavor central charge of the $I_3$ theory.  Also, the CB parameter of an IR free $\U(1)$ gauge theory gives $\D_2 = 1$.  Thus \eqref{kform} gives us that $k=6[2-T({\bf 2h^{(2)}_1})]+T({\bf 2h_1})$.  

Now, we have seen from matching to the $a$ and $c$ central charges from \cite{Chacaltana:2016shw} that $h_1=4$, corresponding to $2h_1=8$ complex scalars (the ``half-hypermultiplets").  8 free half-hypermultiplets can only transform in the ${\bf 2h_1}= 8\cdot {\bf 1}$ (giving $T({\bf 2h_1}) = 0$) or ${\bf 2h_1}= {\bf 4}\oplus\bar{\bf 4}$ (giving $T({\bf 2h_1}) = 2$) representations of an $A_3$ flavor group \cite{McOrist:2013bga}.  In the first case, since the ECB fibers are flavor singlets, they are not lifted under the flavor breaking, but, as singlets, they do not contribute to the index.  In the second case, under the adjoint flavor breaking $A_3\to A_2\oplus\U(1)$ all the ECB fibers are lifted, so $h_1^{(2)}=0$.  So in either case we find $T({\bf 2h^{(2)}_1})=0$ and thus we find from \eqref{kform} that either $k=12$ or $k=14$.  The second is the value found in \cite{Chacaltana:2016shw} from the $\cS$ class construction, and we learn that the ECB fiber transforms in the ${\bf 4}\oplus\bar{\bf 4}$ of the flavor symmetry.

These conclusions also follow from turning on other adjoint breakings of the flavor symmetry.  For example, the minimal adjoint breaking $A_3 \to {A_1}^2\oplus\U(1)$ mass deformation, $m_2$, deforms the singularity as
\begin{align}\label{}
[II^*,A_3\rtimes\Z_2] \xrightarrow{m_2} \left\{ [I_3^*, {A_1}^2\rtimes\Z_2]\, ,\,  [I_1,\U(1)] \right\} .
\end{align}
Since the non-abelian factor appearing in the first $(I^*_3)$ singularity is ${A_1}^2 \simeq D_2 \subset A_3$ with index of embedding $1$, we set $d_1=1$ in \eqref{kform}.  The $I_3^*$ singularity is the IR free gauge theory $\SU(2)$ w/ $4\cdot{\bf 2}\oplus1\cdot{\bf 4}$ massless half-hypermultiplets. The four doublet half-hypermultiplets transform in the $\bf 4$ of the $D_2$ flavor symmetry.  They thus contribute $k_1 = 2\cdot T({\bf 4}) = 4$ to the $D_2$ flavor central charge of the $I_3^*$ theory.  The ECB fibers are either lifted or are flavor singlets.  Thus \eqref{kform} again gives us that $k=12+T({\bf 2h_1})$ (as it must). 

The $\ell=6,4,3$ $\cN=3$ SCFTs all have abelian flavor symmetries, so their central charges cannot be computed by \eqref{kform}.  Note that, since the $\U(1)$ flavor symmetry of these theories is part of the $\cN=3$ $U(3)$ R-symmetry, its central charge is proportional to the $a=c$ central charge.  (The coefficient of proportionality depends on an arbitrary normalization of the $\U(1)$ flavor generator.) 

\subsection{ECB fibers
\label{sec3.3.4}}

The central charge matching performed above showed that the ECB fiber of the $[II^*, A_3\rtimes\Z_2]$ SCFT has complex dimension $2h_1=8$ which transform as ${\bf 4}\oplus\bar{\bf 4}$ under the $A_3$ flavor symmetry.  Also, we saw that the $\ell=6,4,3$ $\cN=3$ theories of each have ECB fiber of complex dimension 2 transforming as $(+1)\oplus(-1)$ under the $\U(1)$ flavor symmetry.  As we now explain, through RG flows we can compute the ECB fiber dimensions of the remaining 4 blue-shaded theories in table \ref{tab1}.

Consider the $[II^*,A_3\rtimes\Z_2]$ theory.  In section \ref{sec3.3.2} we found that the $A_3$ mass deformation $\m_1=\m_2=0$ and $\m_3=-\m_4$ is the one which flows to the $[III^*, A_1\oplus(\U(1)\rtimes\Z_2)]$ theory.   Since the half-hypers of the ECB fiber of the UV theory transform as ${\bf 4}\oplus\bar{\bf 4}$ under the $A_3$ flavor symmetry, upon turning on this adjoint $A_3$ mass two of the four hypermultiplet directions are lifted, leaving unlifted half-hypers in the ${\bf 2}_{+q} \oplus {\bf 2}_{-q}$ of the IR $A_1\oplus(\U(1)\rtimes\Z_2)$ symmetry.  (Here $\pm q$ are the $\U(1)$ charges which may be non-zero because of IR mixing with other global $\U(1)$'s.)  This shows that $h_1=2$ for the $[III^*, A_1\oplus(\U(1)\rtimes\Z_2)]$ theory.  Using this fact together with the central charge formulas \eqref{CCs} and the $III^*\to\{{I_1}^2,I_1^*\}$ deformation pattern gives $a=15/8$ and $c=2$. 

The non-adjoint breaking in \eqref{IIIsA1A1-RG} flows to the $[IV^*,\U(1)]$ theory which, since it is an $\cN=3$ theory, has $h_1=1$.  Under this breaking the 4 half-hypermultiplets of the $[III^*, A_1\oplus(\U(1)\rtimes\Z_2)]$ theory will receive masses $\pm m_1 \pm q m_2 \propto \pm (i \pm q\sqrt3) m_2$, where $q$ is the $\U(1)$ flavor charge of the half-hypermultiplets.  Thus, in order for an ECB fiber hypermultiplet of the $[III^*, A_1\oplus(\U(1)\rtimes\Z_2)]$ theory not to be lifted by this breaking, we must have that $q= i/\sqrt3$ relative to the (arbitrary) normalization of the $\U(1)$ flavor factor chosen by the normalization of the $m_1$ and $m_2$ masses appearing in the SW curve constructed in \cite{Argyres:2015gha}.  (The phase, $i$, in the charge is also arbitrary, since the masses are in the complexified Cartan subalgebra of the flavor symmetry.)  In any case, we learn that the $\U(1)$ flavor charges of the ECB hypermultiplets of the $[III^*, A_1\oplus(\U(1)\rtimes\Z_2)]$ theory are non-vanishing.

The minimal adjoint breaking in the top line of \eqref{A1U1Z2mab} leaves the nonabelian $A_1$ flavor factor unbroken while flowing to lagrangian IR theories.  This breaking can thus be used as in the flavor central charge discussion of section \ref{sec3.3.3} to compute  the $A_1$ flavor factor central charge, $k$, of the $[III^*, A_1 \oplus (\U(1)\rtimes\Z_2)]$ theory.  Since this breaking only turns on a mass for the $\U(1)$ flavor factor, and since we have just learned that the ECB hypermultiplets are charged under this $\U(1)$, it follows that they will be lifted by this flow.  This means that $h_1^{(i)}=0$ in \eqref{kform}, giving $k=10$.

We have, in this way, determined the central charges and ECB fiber dimensions of the $[III^*, A_1 \oplus (\U(1)\rtimes\Z_2)]$ theory shown in table \ref{tab1}.  Perhaps the information on the (pure) Higgs branch structure of the $[II^*,A_3\rtimes\Z_2]$ theory computed in \cite{Chacaltana:2016shw} together with its flow to the $[III^*, A_1\oplus(\U(1)\rtimes\Z_2)]$ theory can be used to also determine the latter's Higgs branch, and, in particular, its quaternionic dimension $h_0$.

Next consider the $[II^*,A_2\rtimes\Z_2]$ theory, which the analysis of section \ref{sec3.3.2} showed might flow to the $[III^*,\U(1)\rtimes\Z_2]$ theory.  If we describe the mass deformations of this theory in terms of $A_2$ masses, $\{m_1,m_2,m_3\}$ with $\sum_i m_i=0$, then the flow to the $[III^*,\U(1)\rtimes\Z_2]$ theory is in the $m_1=0$ direction.  Thus if the $[II^*,A_2\rtimes\Z_2]$ has an ECB fiber transforming in the ${\bf 3}\oplus\bar{\bf 3}$ of $A_2$, one of its hypermultiplets will not be lifted in this breaking, implying (correctly) that the $[III^*,\U(1)\rtimes\Z_2]$ has $h_1=1$ ECB hypermultiplet transforming as $(+1) \oplus (-1)$ under the unbroken $\U(1)$.   It is not hard to see that no other assignment of $A_2$ transformation properties of the $[II^*,A_2\rtimes\Z_2]$ ECB fiber gives the correct result.  

We can use this flow to determine the $a$ and $c$ central charges of the $[II^*,A_2\rtimes\Z_2]$ theory.  Since the singularity splits as $II^* \to \{I_1, III^*\}$ which contribute $N(I_1)=1$ and $N(III^*)=9/2$ (since the $III^*$ singularity must be identified with the $[III^*,\U(1)\rtimes\Z_2]$ theory), \eqref{CCs} gives
\begin{flalign*}
[II^*,A_2\rtimes\Z_2]: &&
a&= \frac{71}{24}, & 
c&= \frac{19}{6}, &
h_1 &= 3 .&&\qquad
\end{flalign*}

If we instead consider the flow where we turn on $A_2$ masses $m_1=m_2$, breaking $A_2 \to A_1 \oplus \U(1)$, the $[II^*,A_2\rtimes\Z_2]$  singularity splits as $II^*\to\{I_2, IV^*_{Q=1}\}$.  This adjoint breaking lifts the ECB fiber so contributes $h^{(i)}_1=0$ and $d_i=1$ on the right side of \eqref{kform}.  The $I_2$ is the IR free $\U(1)$ gauge theory with two charge-1 hypermultiplets, so is the one carrying the unbroken non-abelian $A_1$ flavor factor, and contributes $\D_i=1$ and $k_i=2$ to the right side of \eqref{kform}.  We thus learn that the flavor central charge of the $[II^*,A_2\rtimes\Z_2]$ is $k= 14$.

Finally, consider the flows from the three possible $\ell=4,6$ $\cN=3$ theories to the frozen $[III^*,\varnothing]$, $[IV^*_{q = {\scriptscriptstyle\sqrt2}}, \varnothing]$, and $[IV^*_{q = 1}, \varnothing]$ SCFTs.  In each case the $\U(1)$ mass lifts the ECB fiber and Higgs branch of the $\cN=3$ theory, and so we conclude that $h_0=h_1=0$ for the frozen SCFTs.  (Note that the mechanism described in \cite{Argyres:2012fu} where Higgs branches of SCFTs at the IR end of RG flows are lifted all along the flow does not apply here: because the dimension of the CBs of both the UV and IR SCFTs is the same, there can be no irrelevant gauging of flavor symmetries.)  

\section{RG flow constraints for all flavor assignments
\label{sec4}}

In section \ref{sec2} we pointed out that a given SW curve with discrete symmetry group $\G$ is compatible with multiple choices of the flavor group; see \eqref{misident} and \eqref{misident2}.  In this section we systematically analyze each of these possibilities and discuss which alternative interpretations of the flavor symmetry algebras are allowed.  The main constraint comes from a careful analysis of RG flows and the pattern of factorizations of the curve discriminant. 

Turning on masses breaks the flavor symmetry $\ff$ of the original SCFT.  Since the masses appear as vevs of vector multiplets upon weakly gauging $\ff$, they can be thought of as linear coordinates on $\tf_\C$, the complexified Cartan subalgebra of $\ff$.  Thus the subalgebra of $\ff$ which leaves invariant a given mass deformation is a symmetry of the IR theory.  We call this the \emph{expected} IR flavor symmetry. 

The IR flavor symmetry also manifests itself in the flavor symmetries of the massless degrees of freedom associated to the singularities on the CB.  Mass deformations which leave non-abelian factors of the flavor symmetry unbroken do not fully split the initial singularity.  This is reflected in the occurence of higher-order zeros of the discriminant and correspond to non-frozen conformal or IR-free theories which themselves have unbroken flavor symmetries.  All of these factors will be part of the flavor symmetry in the IR.  We call this the \emph{curve} flavor symmetry which need not be the same as the \emph{expected} IR flavor symmetry.  

We can then distinguish three types of RG flows: \emph{matching} flows are those for which the curve and expected symmetries match; \emph{compatible} flows are those where the expected flavor symmetry is a subalgebra of the curve flavor symmetry of the same rank; and \emph{unphysical} flows are the remainder, i.e., flows for which the curve flavor symmetry either does not contain or is of larger rank than the expected symmetry.  As was argued in \cite{Argyres:2015ffa,Argyres:2015gha}, unphysical flows are indeed unphysical; we will give examples below.  Compatible flows require an accidental enlargement of the flavor symmetry in the IR, while matching ones do not.

This classification of flows gives rise to a classification of the (possible) SCFTs corresponding to the original (UV) singularity from which the flows originate: \emph{good} theories are ones for which all flows are matching; \emph{ugly} theories are ones for which at least one flow is compatible and none are unphysical; and \emph{bad} theories have at least one unphysical flow.  We have no rational reason to exclude ugly theories, only prejudice.

\begin{figure}[tbp]
\centering
\begin{tikzpicture}
[
auto,
good/.style={rectangle,rounded corners,fill=green!50,inner sep=2pt},
bad/.style={rectangle,rounded corners,fill=red!40,inner sep=2pt},
ugly/.style={rectangle,rounded corners,fill=blue!08,inner sep=2pt},
goodarrow/.style={->,shorten >=1pt,very thick,green!70!black},
badarrow/.style={->,shorten >=1pt,very thick,red},
uglyarrow/.style={->,shorten >=1pt,very thick,blue}
]
\begin{scope}[xshift=-0cm]
\begin{scope}[yshift=-0cm]
\node at (3.5,-.5) {{\large{\fontfamily{qcs}\selectfont\textsc{$IV_1^*$ Series}}}};
\end{scope}
\begin{scope}[yshift=-2cm]
\fill[color=yellow!70, rounded corners] (0,.5) rectangle (7,-.5);
\node at (.6,0) {$II^*:$};
\node (G2) at (1.5,0) [good,align=center] {$G_2$};
\node (A2) at (3.5,0) [good,align=center] {$A_2\rtimes\Z_2$};
\node (2U1) at (5.5,0) [ugly,align=center] {$\U(1)^2\rtimes\G_{G_2}$};
\end{scope}
\begin{scope}[yshift=-4cm]
\fill[color=yellow!70, rounded corners] (0,.5) rectangle (7,-.5);
\node at (.6,0) {$III^*:$};
\node (A1) at (2.5,0) [good,align=center] {$A_1$};
\node (U1) at (4.5,0) [good,align=center] {$\U(1)\rtimes\Z_2$};
\end{scope}
\begin{scope}[yshift=-6cm]
\fill[color=yellow!70, rounded corners] (0,.5) rectangle (7,-.5);
\node at (.6,0) {$IV^*:$};
\node (fro) at (3.5,0) [good,align=center] {$\varnothing$};
\end{scope}
\draw[goodarrow] (G2) to (A1);
\draw[badarrow] (G2) to (U1);
\draw[uglyarrow] (A2) to (A1);
\draw[goodarrow] (A2) to (U1);
\draw[uglyarrow] (2U1) to (A1);
\draw[goodarrow] (2U1) to (U1);
\draw[goodarrow] (A1) to (fro);
\draw[goodarrow] (U1) to (fro);
\draw[goodarrow] (A2) to (fro);
\end{scope}
\begin{scope}[xshift=8cm,yshift=-1cm]
\begin{scope}[yshift=-0cm]
\node at (2.5,-.5) {{\large{\fontfamily{qcs}\selectfont\textsc{$IV_{\sqrt{2}}^*$\ \ Series}}}};
\end{scope}
\begin{scope}[yshift=-2cm]
\fill[color=yellow!70, rounded corners] (0,.5) rectangle (5,-.5);
\node at (.6,0) {$II^*:$};
\node (A1) at (1.5,0) [good,align=center] {$A_1$};
\node (U1) at (3.5,0) [good,align=center] {$\U(1)\rtimes\Z_2$};
\end{scope}
\begin{scope}[yshift=-4cm]
\fill[color=yellow!70, rounded corners] (0,.5) rectangle (5,-.5);
\node at (.6,0) {$IV^*:$};
\node (fro) at (2.5,0) [good,align=center] {$\varnothing$};
\end{scope}
\draw[goodarrow] (A1) to (fro);
\draw[goodarrow] (U1) to (fro);
\end{scope}
\end{tikzpicture}
\caption{Green, blue and red arrows label matching, compatible and unphysical RG flows, while green and blue backgrounds indicate ``good" and ``ugly" theories respectively.  There is a $[II^*,\U(1)\rtimes\G_{\!G_2}]\to[I_2,A_1\oplus\U(1)]$ flow which is necessarily only compatible, making the theory ugly.
\label{IVs}}
\end{figure}
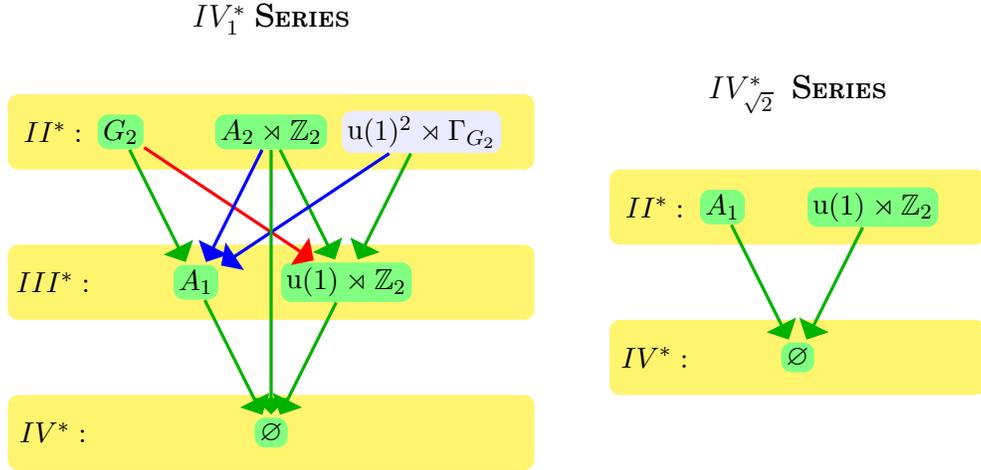

It is a daunting task to algebraically locate all flows which do not fully split a singularity, and classify them as matching, compatible, or unphysical depending on the possible flavor symmetry assignments of the UV and IR singularities.  We are not able to fully perform this task, but instead have examined all minimal adjoint breaking flows and all flows for theories with just two relevant deformations as in \cite{Argyres:2015gha}.  The results are most easily summarized graphically as a web of RG flows among the possible SCFTs which connect different interpretations of the various SW curves.  These are shown in figures \ref{I4}, \ref{I1-I0s}, \ref{IIIs/2I4} and \ref{IVs}, where green, blue, and red arrows denote matching, compatible, and unphysical flows, respectively.  Similarly, theories with a green, blue, or red background are good, ugly, or bad, respectively.  Some compatible flows are not shown in the figures (because they would make them too hard to read) but are explained in the figure captions.

\begin{figure}[tbp]
\centering
\begin{tikzpicture}
[
auto,
good/.style={rectangle,rounded corners,fill=green!50,inner sep=2pt},
bad/.style={rectangle,rounded corners,fill=red!15,inner sep=2pt},
ugly/.style={rectangle,rounded corners,fill=blue!10,inner sep=2pt},
goodarrow/.style={->,shorten >=1pt,very thick,green!70!black},
badarrow/.style={->,shorten >=1pt,very thick,red},
uglyarrow/.style={->,shorten >=1pt,very thick,blue}
]
\begin{scope}[yshift=-0cm]
\node at (7.75,-.5) {{\large {\fontfamily{qcs}\selectfont\textsc{$I_4$ Series}}}};
\end{scope}
\begin{scope}[yshift=-2cm]
\fill[color=yellow!70, rounded corners] (0,.5) rectangle (15.5,-.5);
\node at (.6,0) {$II^*:$};
\node (C5) at (4,0) [good,align=center] {$C_5$};
\node (D5) at (6,0) [ugly,align=center] {$D_5\rtimes\Z_2$};
\node (5A1) at (8.5,0) [ugly,align=center] {${A_1}^5\rtimes S_5$};
\node (5U1) at (11,0) [ugly,align=center] {$\U(1)^5\rtimes \G_{\!BC_5}$};
\end{scope}
\begin{scope}[yshift=-5cm]
\fill[color=yellow!70, rounded corners] (0,1) rectangle (15.5,-1);
\node at (.6,0) {$III^*:$};
\node (C3A1) at (1.5,0) [good,align=center] {$C_3$\\$\oplus$\\ $A_1$};
\node (A3A1) at (3,0) [ugly,align=center] {$A_3\rtimes\Z_2$\\$\oplus$\\ $A_1$};
\node (A3U1) at (5,0) [ugly,align=center] {$A_3\rtimes\Z_2$\\$\oplus$\\ $\U(1)\rtimes\Z_2$};
\node (4A1) at (7,0) [ugly,align=center] {$A_1^3\rtimes S_3$\\$\oplus$\\ $A_1$};
\node (3A1U1) at (9,0) [ugly,align=center] {$A_1^3\rtimes S_3$\\ $\oplus$ \\ $\U(1)\rtimes\Z_2$};
\node (3U1A1) at (11.35,0) [ugly,align=center] {$\U(1)^3\rtimes \G_{\!BC_3}$\\ $\oplus$\\ $A_1$};
\node (4U1) at (14,0) [ugly,align=center] {$\U(1)^3\rtimes \G_{\!BC_3}$\\ $\oplus$\\ $\U(1)\rtimes\Z_2$};
\end{scope}
\begin{scope}[yshift=-8.5cm]
\fill[color=yellow!70, rounded corners] (0,1) rectangle (15.5,-1);
\node at (.6,0) {$IV^*:$};
\node (C2U1) at (4,0) [good,align=center] {$C_2$\\$\oplus$\\ $\U(1)$};
\node (2A1U1) at (7,0) [ugly,align=center] {${A_1}^2\rtimes \Z_2$\\ $\oplus$ \\ $\U(1)$};
\node (3U1) at (11,0) [ugly,align=center] {$\U(1)^2\rtimes \G_{\!BC_2}$\\ $\oplus$\\ $\U(1)$};
\end{scope}
\begin{scope}[yshift=-11.5cm]
\fill[color=yellow!70, rounded corners] (0,.5) rectangle (15.5,-.5);
\node at (.6,0) {$I_0^*:$};
\node (A1) at (5.5,0) [good,align=center] {$A_1$};
\node (U1) at (9,0) [good,align=center] {$\U(1)\rtimes \Z_2$};
\end{scope}
%
\draw[goodarrow] (C5) to [out=200,in=80](C3A1);
\draw[badarrow] (C5) to node [near start,font=\scriptsize] {$\bullet\bullet\bullet$} (A3A1);
\draw[uglyarrow] (D5) to [out=200,in=70] (C3A1);
\draw[goodarrow] (D5) to (A3A1);
\draw[badarrow] (D5) to node [font=\scriptsize] {$\bullet\bullet\bullet$} (A3U1);
\draw[uglyarrow] (5A1) to node [swap,font=\scriptsize] {$\bullet\bullet\bullet$} (4A1);
\draw[goodarrow] (5A1) to (3A1U1);
\draw[badarrow] (5A1) to node [font=\scriptsize] {$\bullet\bullet\bullet$} (3U1A1);
\draw[uglyarrow] (5U1) to node [swap,font=\scriptsize] {$\bullet\bullet\bullet$} (3U1A1);
\draw[goodarrow] (5U1) to (4U1);
\draw[goodarrow] (C3A1) to [out=270,in=140] (C2U1);
\draw[badarrow] (C3A1) to [out=310,in=150] (2A1U1);
\draw[uglyarrow] (A3A1) to (C2U1);
\draw[goodarrow] (A3A1) to [out=300,in=140](2A1U1);
\draw[badarrow] (A3A1) to [out=320,in=160] (3U1);
\draw[uglyarrow] (A3U1) to (C2U1);
\draw[goodarrow] (A3U1) to (2A1U1);
\draw[badarrow] (A3U1) to [out=320,in=145] (3U1);
\draw[uglyarrow] (4A1) to (C2U1);
\draw[goodarrow] (4A1) to (2A1U1);
\draw[badarrow] (4A1) to [out=320,in=130] (3U1);
\draw[uglyarrow] (3A1U1) to (C2U1);
\draw[goodarrow] (3A1U1) to (2A1U1);
\draw[badarrow] (3A1U1) to [out=300,in=110] (3U1);
\draw[uglyarrow] (3U1A1) to (2A1U1);
\draw[goodarrow] (3U1A1) to (3U1);
\draw[uglyarrow] (4U1) to (2A1U1);
\draw[goodarrow] (4U1) to (3U1);
\draw[goodarrow] (C2U1) to (A1);
\draw[badarrow] (C2U1) to (U1);
\draw[goodarrow] (2A1U1) to (A1);
\draw[badarrow] (2A1U1) to (U1);
\draw[uglyarrow] (3U1) to (A1);
\draw[goodarrow] (3U1) to (U1);
\end{tikzpicture}
\caption{Green, blue and red arrows label matching, compatible and unphysical RG flows, while green and blue backgrounds indicate ``good" and ``ugly" theories, respectively.  There are flows of all theories with $\ff=A_1^p\oplus\U(1)^q$ to an $[I_4,A_3\oplus\U(1)]$ theory, with an $\ff\supset A_3$ factor to an $[I_2^*,C_3/A_1^3]$ theory, and the flow $[II^*,D_5\rtimes\Z_2] \to [I_3^*,A_3A_1/A_1^2]$ which render all these theories ugly.
\label{I4}}
\end{figure}
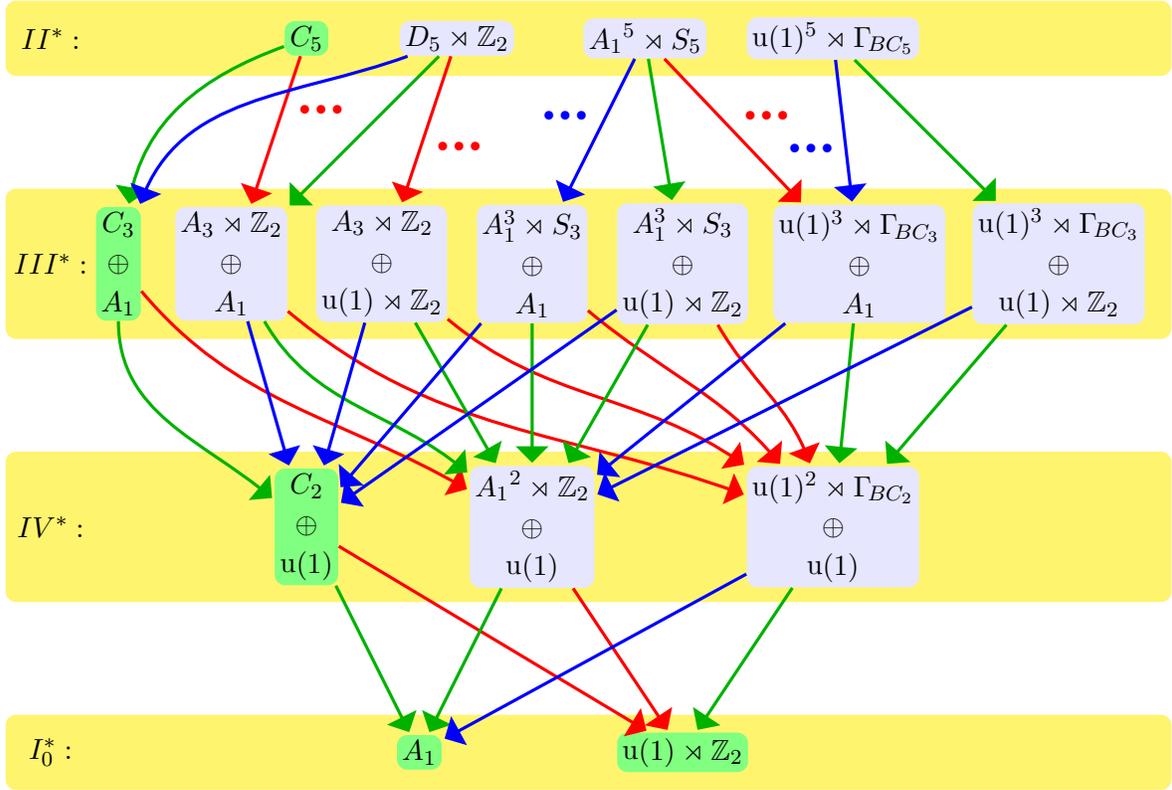

\begin{figure}[tbp]
\centering
\begin{tikzpicture}
[
auto,
good/.style={rectangle,rounded corners,fill=green!50,inner sep=2pt},
bad/.style={rectangle,rounded corners,fill=red!40,inner sep=2pt},
ugly/.style={rectangle,rounded corners,fill=blue!08,inner sep=2pt},
goodarrow/.style={->,shorten >=1pt,very thick,green!70!black},
badarrow/.style={->,shorten >=1pt,very thick,red},
uglyarrow/.style={->,shorten >=1pt,very thick,blue}
]
\begin{scope}[xshift=-0cm]
\begin{scope}[yshift=-0cm]
\node at (2.5,-.5) {{\large{\fontfamily{qcs}\selectfont\textsc{$I_1$ Series}}}};
\end{scope}
\begin{scope}[yshift=-2cm]
\fill[color=yellow!70, rounded corners] (0,.5) rectangle (5,-.5);
\node at (.6,0) {$II^*:$};
\node (E8) at (1.5,0) [good,align=center] {$E_8$};
\node (8U1) at (3.5,0) [ugly,align=center] {$\U(1)^8\rtimes \G_{\!E_8}$};
\end{scope}
\begin{scope}[yshift=-4cm]
\fill[color=yellow!70, rounded corners] (0,.5) rectangle (5,-.5);
\node at (.6,0) {$III^*:$};
\node (E7) at (1.5,0) [good,align=center] {$E_7$};
\node (7U1) at (3.5,0) [ugly,align=center] {$\U(1)^7\rtimes \G_{\!E_7}$};
\end{scope}
\begin{scope}[yshift=-6cm]
\fill[color=yellow!70, rounded corners] (0,.5) rectangle (5,-.5);
\node at (.6,0) {$IV^*:$};
\node (E6) at (1.5,0) [good,align=center] {$E_6$};
\node (6U1) at (3.5,0) [ugly,align=center] {$\U(1)^6\rtimes \G_{\!E_6}$};
\end{scope}
\begin{scope}[yshift=-8cm]
\fill[color=yellow!70, rounded corners] (0,.5) rectangle (5,-.5);
\node at (.6,0) {$I_0^*:$};
\node (D4) at (1.5,0) [good,align=center] {$D_4$};
\node (4U1) at (3.5,0) [ugly,align=center] {$\U(1)^4\rtimes \G_{\!D_4}$};
\end{scope}
\draw[goodarrow] (E8) to [out=225,in=135] (E7);
\draw[goodarrow] (E7) to [out=225,in=135] (E6);
\draw[goodarrow] (E6) to [out=225,in=135] (D4);
\draw[goodarrow] (8U1) to [out=315,in=45] (7U1);
\draw[goodarrow] (7U1) to (6U1);
\draw[goodarrow] (6U1) to (4U1);
\draw[uglyarrow] (8U1) to (E7);
\draw[uglyarrow] (7U1) to (E6);
\draw[uglyarrow] (6U1) to (D4);
\draw[badarrow] (E8) to (7U1);
\draw[badarrow] (E7) to (6U1);
\draw[badarrow] (E6) to (4U1);
\end{scope}
\begin{scope}[xshift=6cm]
\begin{scope}[yshift=-0cm]
\node at (5,-.5) {{\large{\fontfamily{qcs}\selectfont\textsc{$I_1^*$ Series}}}};
\end{scope}
\begin{scope}[yshift=-2cm]
\fill[color=yellow!70, rounded corners] (0,.5) rectangle (10,-.5);
\node at (.6,0) {$II^*:$};
\node (BC3) at (1.8,0) [bad,align=center] {$BC_3$};
\node (A3) at (3.7,0) [good,align=center] {$A_3\rtimes\mathbb{Z}_2$};
\node (3A1) at (6,0) [ugly,align=center] {$A_1^3\rtimes S_3$};
\node (3U1) at (8.5,0) [ugly,align=center] {$\U(1)^3\rtimes \G_{\!BC_3}$};
\end{scope}
\begin{scope}[yshift=-4cm]
\fill[color=yellow!70, rounded corners] (0,.5) rectangle (10,-.5);
\node at (.6,0) {$III^*:$};
\node (2A1) at (2.7,0) [bad,align=center] {$A_1^2$};
\node (A1U1) at (4.9,0) [good,align=center] {$A1\oplus\U(1)\rtimes \Z_2$};
\node (2U1) at (7.8,0) [ugly,align=center] {$(\U(1)\rtimes \Z_2)^2$};
\end{scope}
\begin{scope}[yshift=-6cm]
\fill[color=yellow!70, rounded corners] (0,.5) rectangle (10,-.5);
\node at (.6,0) {$IV^*:$};
\node (U1) at (5,0) [good,align=center] {$\U(1)$};
\end{scope}
\begin{scope}[yshift=-8cm]
\fill[color=yellow!70, rounded corners] (0,.5) rectangle (10,-.5);
\node at (.6,0) {$I_1^*:$};
\node (fro) at (5,0) [good,align=center] {$\varnothing$};
\end{scope}
\draw[goodarrow] (BC3) to [out=280,in=135]  (2A1);
\draw[badarrow] (BC3) to (A1U1);
\draw[badarrow] (BC3) to (2U1);
\draw[uglyarrow] (A3) to (2A1);
\draw[goodarrow] (A3) to (A1U1);
\draw[badarrow] (A3) to  (2U1);
\draw[uglyarrow] (3A1) to (2A1);
\draw[uglyarrow] (3A1) to (A1U1);
\draw[goodarrow] (3A1) to (2U1);
\draw[uglyarrow] (3U1) to (2A1);
\draw[uglyarrow] (3U1) to (A1U1);
\draw[goodarrow] (3U1) to [out=265,in=60] (2U1);
\draw[goodarrow] (2A1) to [out=295,in=140] (U1);
\draw[goodarrow] (A1U1) to (U1);
\draw[goodarrow] (2U1) to [out=245,in=45] (U1);
\draw[goodarrow] (U1) to (fro);
\end{scope}
\end{tikzpicture}
\caption{Green, blue and red arrows label matching, compatible and unphysical RG flows, while green, blue and red backgrounds indicate ``good", ``ugly" and ``bad" theories, respectively.  For the $I_1$ series there is always a compatible flow of any theory with $\ff=\U(1)^p$ to an $[I_5, A_4\oplus\U(1)]$ singularity, rendering them ugly.  For the $I_1^*$ series there are unphysical flows $[II^*,BC_3]\to[I_3^*,A_1^2]$ and $[III^*,A_1^2]\to[I_2^*,\U(1)]$, rendering them ``bad" theories.  The flows of $[II^*,A_1^3\rtimes\Z_2]$ and $[II^*,\U(1)^3\rtimes\G_{\!BC_3}]$ to $[I_3,A_2\oplus\U(1)]$ and from $[III^*,(\U(1)\rtimes\Z_2)^2]$ to $[I_2,A_1\oplus\U(1)]$ are instead only compatible, rendering these theories ``ugly". 
\label{I1-I0s}}
\end{figure}
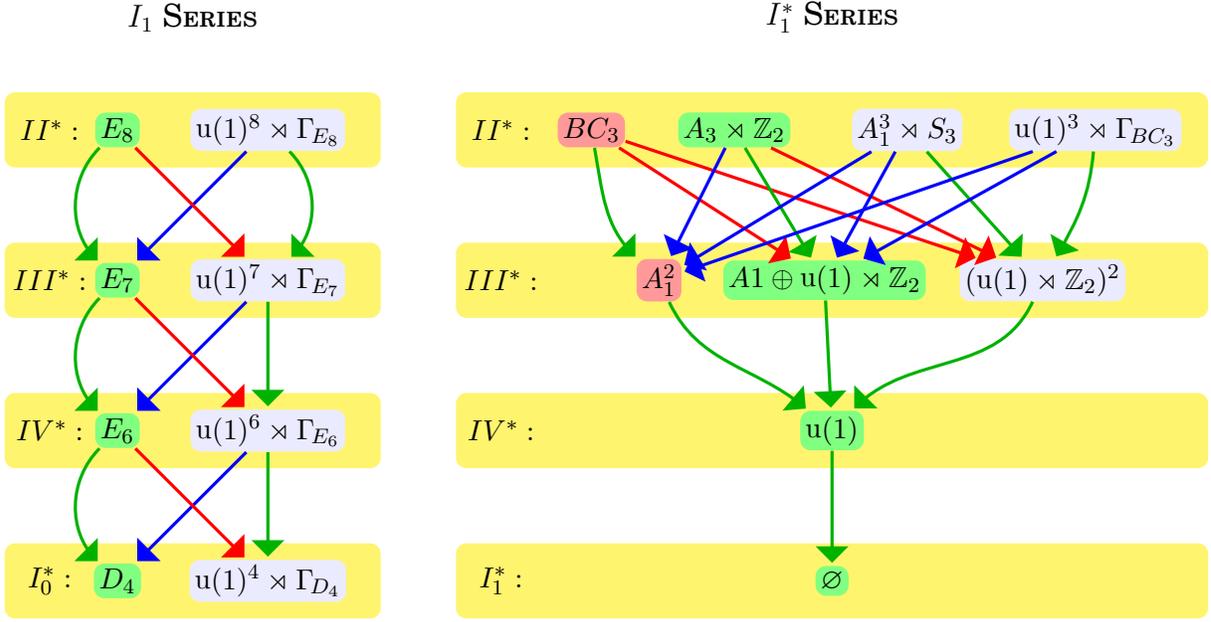

\begin{figure}[tbp]
\centering
\begin{tikzpicture}
[
auto,
good/.style={rectangle,rounded corners,fill=green!50,inner sep=2pt},
bad/.style={rectangle,rounded corners,fill=red!40,inner sep=2pt},
ugly/.style={rectangle,rounded corners,fill=blue!08,inner sep=2pt},
goodarrow/.style={->,shorten >=1pt,very thick,green!70!black},
badarrow/.style={->,shorten >=1pt,very thick,red},
uglyarrow/.style={->,shorten >=1pt,very thick,blue}
]
\begin{scope}[xshift=0cm]
\begin{scope}[yshift=-0cm]
\node at (2.5,-.5) {{\large{\fontfamily{qcs}\selectfont\textsc{$III^*$ Series}}}};
\end{scope}
\begin{scope}[yshift=-2cm]
\fill[color=yellow!70, rounded corners] (0,.5) rectangle (5,-.5);
\node at (.6,0) {$II^*:$};
\node (A1) at (1.5,0) [good,align=center] {$A_1$};
\node (U1) at (3.5,0) [good,align=center] {$\U(1)\rtimes\Z_2$};
\end{scope}
\begin{scope}[yshift=-4cm]
\fill[color=yellow!70, rounded corners] (0,.5) rectangle (5,-.5);
\node at (.6,0) {$III^*:$};
\node (fro) at (2.5,0) [good,align=center] {$\varnothing$};
\end{scope}
\draw[goodarrow] (A1) to (fro);
\draw[goodarrow] (U1) to (fro);
\end{scope}
\begin{scope}[xshift=6cm]
\begin{scope}[yshift=-0cm]
\node at (5,-.5) {{\large{\fontfamily{qcs}\selectfont\textsc{$I_4^2$ Series}}}};
\end{scope}
\begin{scope}[yshift=-2cm]
\fill[color=yellow!70, rounded corners] (0,.5) rectangle (10,-.5);
\node at (.6,0) {$II^*:$};
\node (C2) at (1.8,0) [bad,align=center] {$C_2$};
\node (A2) at (3.7,0) [bad,align=center] {$A_2\rtimes\Z_2$};
\node (2A1) at (6,0) [bad,align=center] {$A_1^2\rtimes S_2$};
\node (2U1) at (8.5,0) [bad,align=center] {$\U(1)^2\rtimes \G_{\!BC_2}$};
\end{scope}
\begin{scope}[yshift=-4cm]
\fill[color=yellow!70, rounded corners] (0,.5) rectangle (10,-.5);
\node at (.6,0) {$I_3^*:$};
\node (2A1b) at (5,0) [good,align=center] {$A_3\oplus A_1$ or higher rank};
\end{scope}
\draw[badarrow] (C2) to (2A1b);
\draw[badarrow] (A2) to (2A1b);
\draw[badarrow] (2A1) to (2A1b);
\draw[badarrow] (2U1) to (2A1b);
\end{scope}
\end{tikzpicture}
\caption{Green and red arrows label matching and unphysical RG flows, while green and red backgrounds indicate ``good" and ``bad" theories, respectively.  
\label{IIIs/2I4}}
\end{figure}
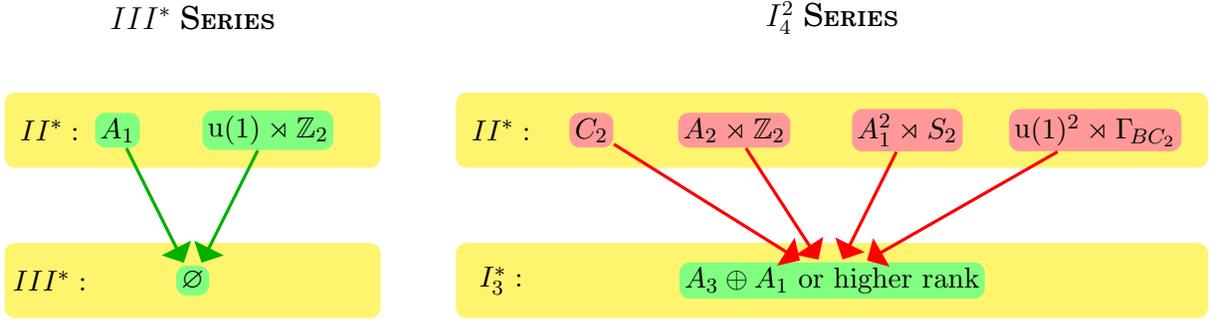

We now illustrate these considerations with a few examples.  First, consider the $[II^*,C_5]$ theory in the $I_4$ series, shown in figure \ref{I4}.  The arrows from the $[II^*,C_5]$ theory represent the minimal adjoint breaking where the adjoint $C_5$ mass breaks $C_5 \to C_3 \oplus \U(1) \oplus A_1$, so $C_3\oplus A_1$ is the \emph{expected} (non-abelian) IR flavor symmetry.  From the curve, one finds that this mass splits the singularity as $II^*\to\{III^*,I_1\}$.  The $I_1$ singularity only has the interpretation as the IR free $\U(1)$ gauge theory with a massless (charge-1) hypermultiplet.  The $III^*$ singularity is some SCFT with a 4-parameter family of relevant deformations.   The seven different possible flavor symmetry interpretations (from the discussion in section \ref{sec2}) of this $III^*$ curve are shown in the $III^*$ row in figure \ref{I4}.  The $[III^*,C_3\oplus A_1]$ curve has a flavor symmetry which matches the expected IR symmetry, and so this flow is a \emph{matching} flow.  In contrast, the $[III^*,(A_3\rtimes\Z_2)\oplus A_1]$ theory has a smaller than expected flavor symmetry, so the flow to it from the $[II^*,C_5]$ theory is unphysical.  The same is true for the remaining flows from the $[II^*,C_5]$ theory to five other theories in the $III^*$ row: this is indicated in figure \ref{I4} by the red dots next to the unphysical flow arrow from the $[II^*,C_5]$ theory. 

Next, consider instead the same minimal adjoint breaking flows but from the $[II^*,{A_1}^5\rtimes S_5]$ theory.   If we label the adjoint masses of the five $A_1$ factors by $m_i$, $i=1,\ldots,5$, then this breaking is given by setting $m_1=m_2=m$ and $m_3=m_4=m_5=0$.  This leaves unbroken a $(\U(1)^2\rtimes\Z_2) \oplus ({A_1}^3\rtimes S_3)$ flavor symmetry --- the expected IR flavor symmetry for this flow.  The singularity of the curve splits as above, and, as above, there is a single flavor interpretation of the $III^*$ singularity which gives a matching flavor symmetry.  Two of the remaining six flavor interpretations of the $III^*$ singularity have smaller-than-expected flavor symmetries, so flows to them are unphysical, while the other four have larger flavor symmetry algebras but with the same expected rank.  Flows to these four are \emph{compatible} flows (shown as blue arrows in the figures):  the accidental IR enlargement of the flavor symmetry is physically allowed, and does not contradict the safely irrelevant conjecture of \cite{Argyres:2015ffa} (which states that there are no $\cN=2$ dangerously irrelevant operators) since the rank of the IR flavor algebra (the number of relevant deformations) is the same as for the expected symmetry.  This illustrates an instance of a general pattern for the minimal adjoint breakings shown in the figures: if one arranges the theories in each row from largest to smallest flavor algebra, then the flows to the left of the matching flow are all compatible, while flows to its right are all unphysical.

Note that for each theory in figure \ref{I4} there is a path of matching flows.  Nevertheless, this does not mean that all the theories have intepretations as ``good" theories.  The reason is that even for interpretations of the minimal adjoint flows shown as matching for these theories, there are other non-minimal flows for which the flavor symmetry does not match.  These flows are not shown in the figure, but are described in the caption.   

In the above examples, because the singularity splits to $\{III^*,I_1\}$, it was easy to figure out the possible IR flavor symmetry assignments since the $I_1$ singularity has a unique interpretation as an IR free theory.  But when the the singularity splits into $I_{n>1}$ or $I_{n>0}^*$ singularities there can be multiple IR-free interpretations of these singularities.\footnote{These IR free interpretations were discussed in detail in \cite{Argyres:2015ffa}.  Other possible, non-lagrangian, interpretations were also discussed in \cite{Argyres:2015gha}, but will not be considered here.}  Often in these cases the number of possibilities can be greatly reduced by looking for consistent interpretations for whole sets of flows.  For example, an $I_1^*$ can be interpreted as an IR free theory as the $\SU(2)$ w/ $10\cdot{\bf 2}$ theory with $\ff=D_5$, or as $\SU(2)$ w/ $2\cdot{\bf 2}\oplus2\cdot{\bf 3}$ with $\ff=\U(1)\oplus A_1$, or as $\SU(2)$ w/ $4\cdot{\bf 3}$ with $\ff=C_2$ (with charge normalization $a=1/2$), or as $\SU(2)$ w/ $1\cdot{\bf 4}$ with $\ff=\varnothing$.  But if there were another mass which further split $I_1^*\to\{{I_1}^7\}$, then only the first, $\ff=D_5$, assignment would be consistent.  But usually we need to account for a large web of possibilities and seek a pattern of matching or compatible RG flows. 

Three final notes on the figures.  First, we have introduced a compact notation $\G_X$ for the Weyl group of the Lie algebra with Dynkin name $X$.  Second, it is important to note that the ``bad" $[II^*,BC_3]$ and $[III^*,{A_1}^2]$ theories in the $I_1^*$ series become ``good" theories if the frozen $I_1^*$ singularity is interpreted as a non-lagrangian field theory as discussed in \cite{Argyres:2015gha}. Finally, we have not analyzed the RG flows for the $I_0^*$ series here since it will be the subject of \cite{Argyres:2016I0s}


\section{Conclusion}

Our main result is to provide evidence for the existence of at least an extra 8 rank 1 4d $\cN=2$ SCFTs in addition to the 11 already known.  Four of them were recently discussed in \cite{Garcia-Etxebarria:2015wns,Chacaltana:2016shw}.  Here we not only point out that they fit into our classification of rank 1 $\cN=2$ SCFTs, but also that their existence implies the existence of additional rank 1 theories through RG flow consistency arguments.  Furthermore, using the techniques developed in \cite{Argyres:2015ffa,Argyres:2015gha,Argyres:2015ccharges} we are also able to further characterize the central charges, ECB fibers, and RG flows of the recently proposed theories.

Technically, we lift an implicit assumption made in \cite{Argyres:2015gha} that flavor symmetries of $\cN=2$ SCFTs have no discrete factors.  Lifting this assumption effectively allows multiple different flavor symmetry interpretations of each CB geometry found in \cite{Argyres:2015gha}.  We have characterized here precisely what the freedom in flavor interpretations is, and have presented a discussion of all the allowed possibilities, summarized in table \ref{tab1} and especially figures \ref{I4}--\ref{IVs}.  

Most notably, the new interpretation of the flavor symmetries of some of the theories in the $I_1^*$ series has ``rehabilitated" the lagrangian interpretation of the frozen $I_1^*$.  That is, the undeformable $I_1^*$ which appears in the deformation patterns of these theories can simply be interpreted as an $\SU(2)$ w/ $1\cdot{\bf4}$ lagrangian theory \cite{Argyres:2015ffa} and not as a non-lagrangian, weakly gauged rank-0 theory, $X_1$, as proposed in \cite{Argyres:2015gha}.  A similar but more subtle story holds for the $I_0^*$ series and will be the subject of \cite{Argyres:2016I0s}.

We believe that being able to systematically discuss the set of possible $\cN=2$ SCFTs which could appear at rank 1 is a remarkable result.  Our findings show that despite decades of continuous advances in our understanding of $\cN=2$ SCFTs, the landscape even of rank 1 theories is not well understood.  Other systematic explorations of the landscape of low-rank $\cN=2$ SCFTs using techniques such as the bootstrap \cite{Beem:2014zpa,Liendo:2015ofa,Lemos:2015awa,Lemos:2015orc}, $\cS$-class constructions \cite{Chacaltana:2010ks,Chacaltana:2011ze,Chacaltana:2012ch,Chacaltana:2013oka,Chacaltana:2014jba,Chacaltana:2014nya,Chacaltana:2015bna,Wang:2015mra,Chacaltana:2016shw}, geometric engineering \cite{DelZotto:2015rca,Xie:2015rpa,Xie:2015xva}, BPS quivers \cite{Alim:2011kw,Cecotti:2012va,Cecotti:2012jx,Cecotti:2013lda,Cecotti:2013sza,Cecotti:2014zga,Cordova:2015nma}, and clarifying and generalizing the F-theory construction of \cite{Garcia-Etxebarria:2015wns} will undoubtably help sharpen our understanding.

\acknowledgments

It is a pleasure to thank M. Del Zotto, J. Distler, C. Long, D. Morrison, L. Rastelli, Y. Tachikawa, Y. Wang, and D. Xie for helpful comments and discussions. This work was supported in part by DOE grant DE-SC0011784.  MM was also partially supported by NSF grant PHY-1151392.

\bibliographystyle{JHEP}
\bibliography{SCFTs_2a}

\end{document}